\documentclass[10pt,preprint2]{aastex}

\usepackage{rotating}

\begin{document}

\title{Triggered Star Formation in Galaxy Pairs at $z=0.08-0.38$
\\ (Short title: ``Triggered Star Formation'')}
 
\author{Deborah Freedman Woods}
\affil{Department of Astronomy, Harvard University, Cambridge, MA 02138}
\email{dwoods@cfa.harvard.edu}

\author{Margaret J. Geller,  Michael J. Kurtz, \\ Eduard Westra, Daniel G. Fabricant }
\affil{Smithsonian Astrophysical Observatory, Cambridge, MA 02138}

\author{Ian Dell'Antonio}
\affil{Department of Physics, Brown University, Providence, RI 02912}

\begin{abstract}

We measure the strength, frequency, and timescale of tidally triggered star formation 
at redshift $z=0.08-0.38$ in a spectroscopically complete sample of galaxy pairs
drawn from the magnitude-limited redshift survey of 9,825 
Smithsonian Hectospec Lensing Survey (SHELS) galaxies with $R<20.3$.   
To examine the evidence for tidal triggering, we identify a volume-limited sample of 
major ($\left | \Delta M_R \right |<1.75$, corresponding to mass ratio $>1/5$) pair 
galaxies with $M_R < -20.8$ in the redshift range $z=0.08-0.31$.  The size and 
completeness of the spectroscopic survey allows us to focus on regions of low 
local density.  The spectrophotometric calibration enables the use of the
the 4000~\AA\ break ($D_n4000$), the H$\alpha$ specific star formation  rate 
(SSFR$_{H\alpha}$), and population models to characterize the galaxies.
  We show that $D_n4000$ is a useful population classification tool; it closely 
tracks the identification of emission line galaxies.  The sample of major pair galaxies 
in regions of low local density with low $D_n4000$ demonstrates the expected 
anti-correlation between pair-wise projected separation and a set of star formation 
indicators explored in previous studies.  We measure the frequency of 
triggered star formation by comparing the SSFR$_{H\alpha}$ in the volume-limited 
sample in regions of low local density:  $32\pm7\%$ of the major pair 
galaxies have SSFR$_{H\alpha}$ at least double the median rate of the 
unpaired field galaxies.   Comparison of stellar population models for 
pair and for unpaired field galaxies implies a timescale for triggered star 
formation of $\sim 300-400$~Myr.  
 \\ \\

\end{abstract}

\keywords{galaxies: interactions -- galaxies: stellar content -- galaxies: active}

\section{Introduction} \label{intro}

Hierarchical galaxy formation models predict frequent interactions in a galaxy's
history \citep[e.g.][]{cole00,wechsler02}.  Some of these interactions may
trigger star formation.  Observational limits on the frequency and duration of
triggered star formation are an important constraint on the role of these
interactions in the evolution of stellar populations.

Evidence that galaxy interactions at low redshift trigger star formation 
was first detected in photometric observations by \citet{larson_tinsley}, who 
demonstrated that apparently interacting galaxies in the Atlas of Peculiar Galaxies 
\citep{arp66} have a broader range of colors and tidal features than typical galaxies.  
 Spectroscopic indicators confirm that interacting galaxies tend 
to have enhanced star formation rates \citep{kk84,kenn87,keel93,liu-kenn95a,donzelli-p97,
bgk,bgk03,lambas03,nikolic04,kauff04_713,woods06,woods07,ellison08}.
Infrared observations \citep{kenn87,jonesstein89,seg-wol92,keel93,nikolic04,geller06,smith07} 
and radio observations  \citep{hummel81} yield similar results. 
Observations that  galaxy pairs with the smallest project separation have the 
highest star formation rates provide direct evidence for triggered star formation
\citep[e.g][]{bgk}.

Recent studies show that galaxy interactions at intermediate redshift trigger star 
formation.  Measurements of infrared luminosity  \citep{lin07} and 
EW([O II]) \citep{deravel08} indicate an anti-correlation between projected 
separation and star formation activity for pair galaxies at intermediate
 redshift ($z < 1$), similar to the trend observed at low redshift.  
Analysis of the morphology of  intermediate redshift ($z=0.1-0.6$)
pair galaxies likewise shows an increase in the fraction of asymmetric 
galaxies in pairs, an indication of tidal interaction \citep{patton05}.

Although observations demonstrate that galaxy interactions {\it can} 
lead to enhanced star formation activity, the results of studies of the frequency 
and intensity of star formation activity appear to vary with the sample selection.
 \citet{li08}  find that $30\%$ of the SDSS pair galaxies with high specific star 
formation rates have a companion within 100~kpc, and that
low to average specific star formation rate galaxies rarely have a companion.
In a separate study of SDSS pair galaxies at $z<0.16$, \citet{ellison08} measure 
up to $40\%$ enhancement in the median star formation rate of close pair galaxies 
 with stellar mass ratios $< 1/10$ relative to their comparison sample of non-pair systems. 
 \citet{bergvall03}  find that global $UBV$ colors
do not show significant enhancement in star formation in their small sample of
 interacting galaxies compared to isolated systems.
However, they do find that star formation rates at the very centers of
the interacting systems are increased by a factor of $\sim 2-3$ over 
the non-interacting systems.  Spitzer infrared observations of 35 tidally 
disturbed Arp galaxies show that their $24~\mu$m emission is more
centrally concentrated than in normal spiral galaxies, suggesting a build up
of central gas which could fuel central star formation; their infrared colors
suggest an increase in the mass normalized SFR by a factor of two over the
normal spirals \citep{smith07}.

Numerical simulations account for the observed range of strength, frequency, and 
duration of triggered star formation. In the large suite of
simulations by \citet{dimatteo08}, galaxy interactions and mergers 
produce only moderate star formation enhancement relative to non-interacting galaxies at 
low redshift.  Strong starbursts are rare and short lived, typically lasting a 
few hundred Myr.  The duration of the starburst declines as the enhancement in star 
formation rate increases.

Global galaxy properties affect their susceptibility to tidally triggered star formation.
Systems with sufficient gas can exhibit triggered star formation in major 
interactions \citep[e.g.][]{mihos+hern96,tissera,cox06,dimatteo07}.
 Interactions between gas-poor galaxies (``dry mergers'') produce little or no star 
formation, although they contribute substantially to the build-up of massive 
galaxies  \citep[e.g.][]{tran05,vandokkum05,cattaneo08}.  Internal structure 
strongly affects the susceptibility  to gaseous inflows; late-type
galaxies without strong bulge components are more likely to have
bar instabilities that drive the gaseous inflows 
powering the central star formation \citep{mihos+hern96,dimatteo07}.

The relative mass of the galaxy compared to its neighbor and the intrinsic
mass also influence the degree of triggered star formation.
Galaxy pairs of similar mass (luminosity) show strong enhancement
to their star formation rates in both members of the pair in both observational 
\citep{woods07,ellison08} and theoretial \citep{cox08} studies.
Observations show that although satellite galaxies occasionally experience 
triggered star formation, the brighter of a high mass (luminosity) ratio pair does 
not undergo significant tidal triggering \citep{woods07}; numerical simulations
suggest a similar picture \citep{cox08}.  The intrinsic mass of 
the galaxy also plays a role:
 low mass (luminosity) galaxies in major mergers are more strongly 
affected by the interaction than are high mass galaxies in major mergers in
observations of pair galaxies \citep{woods07,ellison08}.

Local environment can  affect star formation activity in a galaxy apparently
independent of an interaction.  Galaxies in clusters and large groups tend to
have redder color and less current star formation than isolated systems
\citep[e.g.][]{hubble31,cooper,gerke}.  \citet{barton07} show that pair galaxies are
more common in regions with greater local density than are unpaired
galaxies.  Including pair galaxies from high density regions suppresses the
signal of triggered star formation measured in comparison to unpaired galaxies.

We take advantage of the size and completeness of the Smithsonian Hectospec Lensing 
Survey \citep[SHELS,][]{geller05,geller10} galaxy sample to 
identify a set of major interactions at 
$z=0.08-0.38$ for galaxies with young stellar populations in regions of low 
local density, where we can measure the cleanest signal of triggered star 
formation. Local density selection at intermediate redshift has not been possible 
before this sample.  

The stability of the Hectospec instrument enable accurate 
spectrophotometric calibration of the spectra.  The calibrated spectra provide
the 4000~\AA\ break ($D_n4000$) and the H$\alpha$ specific star formation  rate 
(SSFR$_{H\alpha}$), which we use  to characterize the star formation.
We quantify the frequency, strength, and timescale of triggered star formation
in major galaxy interactions using measurement of specific star formation rates and 
stellar population models fitted to the individual spectra.

We describe the SHELS data in \S\ref{data} and the sample selection in \S\ref{sample}.
Section~\ref{properties} describes the classification of galaxies as starforming or 
AGN and the use of the spectroscopic indicator $D_n4000$ to identify galaxy types.  
We  consider local density effects in \S\ref{density}.
Section~\ref{evidence} describes various measurements of tidally triggered star formation,
including the frequency, strength, and timescale of enhanced star formation activity.
 Throughout this study we assume standard $\Lambda$CDM cosmology, where 
$H_0 = 71$~km~s$^{-1}$~Mpc$^{-1}$, $\Omega_m = 0.3$, and $\Omega_{\Lambda} = 0.7$
 \citep{wmap}.

\section{Data} \label{data}

In this section we describe the Smithsonian Hectospec Lensing Survey 
data set (SHELS).  We give an overview of the data set in \S\ref{overview}.  
We describe the photometric measurements in \S\ref{phot}, and the spectroscopic
measurements in \S\ref{specmeas}, including the H$\alpha$ specific
star formation rate in \S\ref{sfr}, and aperture effects in \S\ref{aperture}.

\subsection{Smithsonian Hectospec Lensing Survey} \label{overview}

The SHELS magnitude-limited redshift survey includes 9,825 galaxies with total 
$R$-band magnitude $R<20.3$.
The galaxies are selected from the Deep Lens Survey F2 field  R-band images 
\citep[DLS;][]{wittman02,wittman06}.  We obtained spectroscopy using Hectospec 
\citep{fabricant05} at the MMT.   The spectroscopic sample is $97.7\%$ 
complete to R=20.3, and the differential completeness at the magnitude
limit is $94.6\%$. 

We restrict our analysis to the 6,935 galaxies with redshift 
$z= 0.080-0.376$.  The lower redshift limit minimizes aperture effects
(\S\ref{aperture}) and the upper limit allows measurement of the H$\alpha$ flux
(\S\ref{specmeas}).

The DLS F2 field covers 4 deg$^2$ on the sky centered on 
R.A. = $9^{\mbox{\small h}}19^{\mbox{\small m}}32_{.}^{\mbox{s}}4$, 
decl.=+30$^{\circ}$00\arcmin00\arcsec\ (J2000).  The Sloan Digital Sky Survey 
\citep[SDSS;][]{sdss} also covers this region. 

The DLS images come from the Kitt Peak Mayall 4-meter telescope with the 
 Mosaic prime-focus imager (Muller et al. 1998).  The DLS  observed in the 
$R$-band during nights with seeing $\leq 0$\farcs9.  The 1-$\sigma$ limiting surface 
brightness in $R$-band is 29 mag~arcsec$^{-2}$.  The DLS also observed in $V$-band. 

The 0\farcs9 resolution of the DLS $R$-band images  allows identification of 
close pairs or merging galaxies with minimum separations of $15-20$~kpc up to 
$z \leq 0.3$.    The median seeing for the SDSS photometry is 
1\farcs4 PSF in $r$-band \citep{sdss}.  Thus some systems that cannot be
distinguished in the SDSS data can be identified in the DLS.  A total of eight 
pairs in the DLS are missing from SDSS.
 
Objects with stellar light profiles, i.e. AGN at high redshift, 
may be preferentially excluded from our sample because DLS objects with stellar 
light profiles are not targeted for spectroscopy.  However, with the DLS 
29~mag~arcsec$^{-2}$
surface brightness limit and the 0\farcs9 resolution, we should be able to detect
bright host galaxy bulges to the limiting $z=0.376$.  We discuss AGN detection
in more detail in \S\ref{agn}.

The Hectospec observations identify a spectroscopic pair of emission
line galaxies at R.A.= $9^{\mbox{\small h}}16^{\mbox{\small m}}58_{.}^{\mbox{s}}903$, 
decl. = +29$^{\circ}$43\arcmin10\farcs597 (J2000).   The image,
which shows a double-nuclei extended object, and the spectra are available on 
our website\footnote{http://tdc-www.cfa.harvard.edu/instruments/hectospec/progs/tsf/}. 
We do not include this object in our analysis  because the selection of spectroscopic 
pairs is not complete or uniform. 

Only 230 galaxies in our sample lack spectra ($2.3\%$). 
Of these, $\sim 24$ galaxies ($10\%$) may be members of pairs that 
satisfy our criteria for analysis (\S\ref{sample}).  The fraction of 
galaxies in pairs in our sample as a whole is $\sim10\%$ (\S\ref{sample}).

\subsection{Photometric measurements} \label{phot}

 Image processing for the DLS uses the IRAF package {\tt MSCRED} to correct for 
 bias, to  flat field, and to obtain basic astrometric calibration. \citet{wittman06}
construct a stacked DLS
image of each subfield in the $R$-band, correcting for cosmic rays and saturated 
objects, fixing the astrometric calibration, and correcting the image shape and 
photometry for optical distortion.  The uncertainty in the
$R$-band magnitude is $\sim 0.05$~mag.  We derive the galaxy
catalog using SExtractor \citep{sextract} on the final stacked
images.  We remove objects within the radius of the diffraction pattern 
around bright stars.  The total area excluded around 778 stars is 
0.215~deg$^2$ ($5.4\%$ of the survey).  

The absolute magnitude $M_R$ calculation includes both k- and
evolutionary  (e) corrections.  We use the k+e correction determined
by \citet{annis01} using the Pegase code \citep{pegase}.  The k+e 
correction requires classification as one of nine galaxy types
(bright cluster galaxy, elliptical, S0, Sa, Sb, Sbc, Sc, Sd, and irregular).
We classify the galaxies according to the SDSS ($g-i$) color and galaxy
redshift. We apply the SDSS $r$-band k+e correction
to the DLS $R$-band photometry because the shape of the filter band passes 
are similar and the difference in corrections for the different 
bands is negligible.  We estimate the uncertainty in the k+e correction 
from the difference in the correction  for adjacent galaxy types.  
The uncertainty ranges from a maximum of $\sim0.02$ mag at $z=0.08$ 
to  0.03 -- 0.13~mag at $z=0.38$, depending on the galaxy
type.

 There are 51 galaxies in our $z=0.080-0.376$ sample with a Hectospec redshift 
but no SDSS photometry for galaxy classification and k+e correction, including 
three galaxies in major pairs (\S\ref{majpair}).  In 34 of the 51 cases, we 
use the DLS ($V-R$) color to estimate the ($g-i$) color  for the k+e correction.  For the 
17 cases where we lack both DLS ($V-R$) and SDSS data (including two in major
pairs), we assume the galaxy type is Sa.  The k+e correction for the galaxy type 
Sa is in the middle of the range for the different galaxy types; choosing a 
different k+e correction has no significant effect on the results.  
At the redshift $z=0.3$, the apparent $R=20$.3 limit corresponds 
to a k+e corrected absolute magnitude $M_R = -21.4$ for a typical late-type 
galaxy, and to $M_R =-21.8$ for a typical early-type galaxy.

\subsection{Spectroscopic Measurements} \label{specmeas}

The Hectospec spectrograph  270 line mm$^{-1}$ grating yields $\sim 
6$~\AA~pixel$^{-1}$ dispersion over the wavelength range 3700-9300~\AA~ 
\citep{fabricant08}.  The 300 optical fibers with 1\farcs5 diameter are placed 
radially within a 1$^{\circ}$ diameter field.  About 30 of 
the fibers measure sky background during each pointing.

We reduce Hectospec data with the standard Harvard-Smithsonian Center for 
Astrophysics  Hectospec reduction 
pipeline\footnote{http://tdc-www.harvard.edu/instruments/hectospec/reduce.html}
\citep{fabricant05,mink07}. We measure the galaxy redshift using the program {\tt RVSAO} 
\citep{kurtz_mink98,fabricant05}.  The total RMS internal error in the redshift is 
34 km/s for emission line galaxies and 65 km/s for absorption line galaxies,
based on repeat observations of 812 emission lines and 542 absorption line galaxies.
We further process spectra  to correct for atmospheric extinction
and to remove the narrow absorption lines from H$_2$O and O$_2$ at wavelengths
longer than $\sim 6000$~\AA.  We also  correct for relative throughput as a
function of wavelength \citep{fabricant08}.

We compute the emission line flux by integrating the counts in a fixed-width 
band in the rest-frame  centered on the central wavelength of the line;  we 
 subtract the continuum level.  The spectral regions where we measure H$\alpha$ and
H$\beta$ emission are $6562.8 \pm8$~\AA\ and $4861.3 \pm8$~\AA, respectively.
We compute the continuum  from the average of 
regions on either side of the band after rejecting noisy data.
The RMS scatter in the line flux is $23\%$ for H$\alpha$, and 
internal systematic error in H$\alpha$ is $\sim18\%$, based on duplicate 
observations of 592 galaxies.  

The calibrated $R$-band photometry from the DLS in a 1\farcs5 diameter 
aperture serves as reference for conversion of counts to flux
in the 1\farcs5 diameter fiber.  This calibration process removes variation
caused by seeing, clouds, telescope tracking and guiding, and astrometry
and alignment errors for  light losses that are wavelength
independent \citep{fabricant08}.

Comparison of Hectospec spectra with a sample of overlapping SDSS spectra 
shows that the median ratio of the spectra is constant to $\sim5\%$ over the 
range $3850-8000$~\AA, and the offsets in the ratios of the median
H$\alpha$ flux and median rest-frame H$\alpha$ equivalent width (EW(H$\alpha$)) 
are $\sim3\%$ \citep{fabricant08}.

We also compute the spectroscopic indicator $D_n4000$,  a measure of the 
stellar population age, which is defined as the ratio of flux in 
the band 4000-4100~\AA~ to flux in 3850-3950~\AA \citep{balogh99}.  
The RMS scatter in $D_n4000$ is 0.086.  The 
internal systematic errors in  $D_n4000$ are very small, $4.5\%$,
based on repeat measurements of 1468 galaxies.  The comparison of our
measured $D_n4000$ with an overlapping sample of SDSS spectra shows a median 
ratio of 1.00 \citep{fabricant08}.

\subsubsection{H$\alpha$ specific star formation rate} \label{sfr}

We compute the H$\alpha$ specific star formation rate (SSFR$_{H\alpha}$) 
from the  H$\alpha$ flux ($f_{H\alpha})$ normalized by the $R$-band 
1\farcs5 aperture magnitude.  The H$\alpha$ flux must first be 
corrected for stellar absorption and for reddening. 

 To determine the stellar absorption at H$\alpha$ and H$\beta$,
 we fit the individual spectrum using the \citet{tremonti04}
continuum subtraction method (see also \citealp{westra10}).
This method fits the calibrated spectra with stellar population 
models to estimate the ages of the stellar populations within the galaxy.
Like \citeauthor{tremonti04}, we assume that the star formation history of a galaxy 
can be represented as the sum of discrete bursts of star formation, and 
we fit for the burst ages.

\begin{figure}[htb]
  \begin{center}
 \includegraphics[width=2.25in,angle=90]{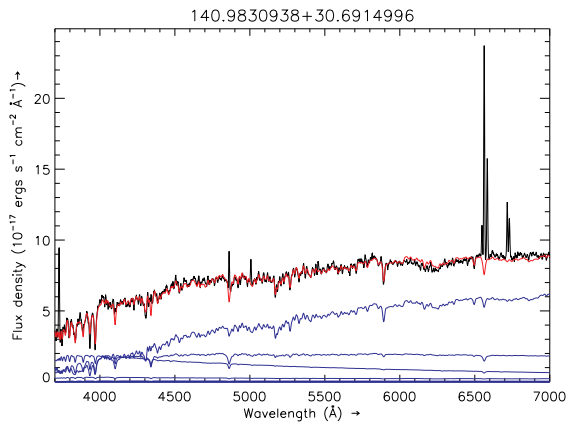}
 \caption{An example rest-frame spectrum illustrating the contribution of
 different stellar populations to the continuum  galaxy spectrum.
The solid black line is the observed galaxy spectrum.  
The solid red line shows the modeled galaxy continuum from the sum of the 
discrete starburst populations, shown as dashed and dash-dot lines (blue).}
 \label{newfig}
 \end{center}
\end{figure}

Figure~\ref{newfig} illustrates the fit of the contributions from the
different starburst populations in an example galaxy spectrum.
 The template spectra that we use are \citet{bruzual03}
stellar population models. The models include 10 different ages of
bursts (0.005, 0.025, 0.1, 0.3, 0.6, 0.9, 1.4, 2.5, 5 and 10 Gyr) at
Solar metallicity.    We also add reddening as an additional 
parameter using the \citet{charlot00} attenuation model. 
 Finally, we match the models  in redshift, pixel size, and spectral resolution 
to each galaxy spectrum.  The core of the code adapted from \citeauthor{tremonti04} 
uses the IDL {\tt mpfitfun} routine from the Markwardt 
library \citep{mpfit} to find the best fitting combination of models.

The absorption-corrected H$\alpha$ flux is:
\begin{equation} \label{abscorr}
f_{c,\mbox{H}\alpha} = f_{H\alpha} \left ( \frac{EW(H\alpha) + EW_{Abs}}{EW(H\alpha)} \right ),
\end{equation}
and similarly for $f_{c}(\mbox{H}\beta)$. The ratio of $f_c/f$ is equivalent 
to $EW_c/EW$.  All EW and line fluxes of H$\alpha$ or H$\beta$ 
in this paper are absorption corrected.

We correct the H$\alpha$ flux for reddening  according to the standard
method  \citep{calzetti00}. We compute  the wavelength dependent 
extinction,  $A_{\lambda}$, where
the factors $k(H\alpha)=3.326$ and $k(H\beta)=4.598$ give the differential 
extinction between the wavelengths of $H\alpha$ and H$\beta$ for the 
case of a starburst galaxy \citep[$k(V) = 4.05$,][]{calzetti00}.
The parameter $R_{\alpha\beta}$ relates the observed to intrinsic Balmer 
decrement \citep[2.87;][]{calzetti01}.

\begin{figure}[htb]
  \begin{center}
 \includegraphics[width=2.25in,angle=90]{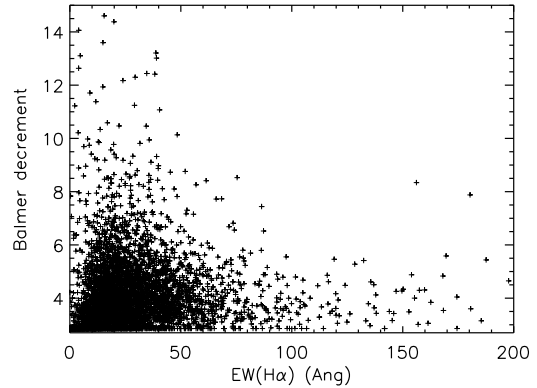}
 \caption{EW(H$\alpha$) versus Balmer decrement for all emission line galaxies. 
 The galaxies with EW(H$\alpha) < 20$~\AA~ and Balmer decrement $> 10$  result from
noisy H$\beta$ measurements.  We use the median Balmer decrement in bins of 
absolute luminosity and redshift.  We set all Balmer decrement
$< 2.86$ to that value.  Measurement uncertainty in EW(H$\alpha$) is
$18\%$ and the uncertainty in the Balmer decrement is $25\%$.}
 \label{balmer}
 \end{center}
\end{figure}

We correct for reddening using average  $E(B-V)$ values for bins of absolute 
luminosity and redshift.  This procedure mitigates the effects of noisy individual 
H$\beta$ measurements, which contribute significant noise to the  $E(B-V)$ values. 
Figure~\ref{balmer} shows the EW(H$\alpha$) versus  Balmer decrement for individual
galaxies.  A population of galaxies with EW(H$\alpha) < 20$~\AA~
and Balmer decrement $> 10$ result from noisy H$\beta$ measurements.
The total internal error in EW(H$\alpha$) is $18\%$ and the error in the Balmer
 decrement is $25\%$.  We determine the Balmer decrement for each luminosity and 
redshift bin from the median value in the appropriate bin derived from the 484 
galaxies with S/N$\geq3$ in H$\alpha$, S/N$\geq2$ in H$\beta$, EW(H$\alpha) > 3$~\AA,
EW(H$\beta) > 2$~\AA, $D_n4000 \leq 1.44$, and classified as starforming (\S\ref{agn}).
Using the median $E(B-V)$ value  eliminates huge and unphysical reddening correction 
factors for individual galaxies. 

Once we correct the H$\alpha$ flux 
for stellar absorption and for reddening, we calculate the H$\alpha$ luminosity 
($L_{H\alpha}$) using the luminosity distance.  We convert from $L_{H\alpha}$ to 
star formation rate using the conversion factor of \citet{kennicutt98}:
\begin{equation}
SFR_{H\alpha}(M_{\sun} \mbox{ yr}^{-1}) = \frac{L_{H\alpha}}{1.27\times10^{41} \mbox{erg s}^{-1}}.
\end{equation}

We normalize the H$\alpha$ star formation rate by the galaxy $R$-band aperture
magnitude to compute the  H$\alpha$ specific star formation rate (SSFR$_{H\alpha}$).
 We use the 1\farcs5 diameter $R$-band aperture magnitude 
from the DLS to match the fiber diameter where the H$\alpha$ flux is measured.  
Thus, no correction for aperture covering fraction is needed.
The normalization factor assumes
$1 M_{\sun} = 1 L_{\sun}$ in the $R$-band and $M_{R(\sun)} = 4.42$ 
\citep{binney}. 

The uncertainty in the SSFR$_{H\alpha}$ follows from the $23\%$ RMS error
in the H$\alpha$ flux and the 0.05~mag uncertainty in the normalization.  The
absorption and reddening corrections contribute additional uncertainty of 
$\sim 20\%$.  We add these contributions to the error in quadrature to
estimate a total uncertainty of $\sim30\%$ in estimates of  SSFR$_{H\alpha}$.

\subsubsection{Aperture effects} \label{aperture}

The 1\farcs5 Hectospec aperture implies that the aperture covering
fraction varies as a function of galaxy luminosity and redshift. \citet{kewley05}
suggest that aperture covering fractions  greater than
$20\%$ yield star formation rates  representative of the galaxy 
as a whole.  The aperture size required to meet this standard varies with both
redshift and absolute luminosity because brighter galaxies are generally larger
\citep{kewley05}. In the volume-limited sample (defined in \S\ref{majpair}), 
the galaxy radius included in the 
fiber varies from $r = 1.1$~kpc at the lower redshift limit ($z=0.080$) to 
$r = 3.4$~kpc at the upper redshift limit of the volume-limited sample 
($z=0.310$).  

\begin{figure}[htb]
  \begin{center}
 \includegraphics[width=2.15in,angle=90]{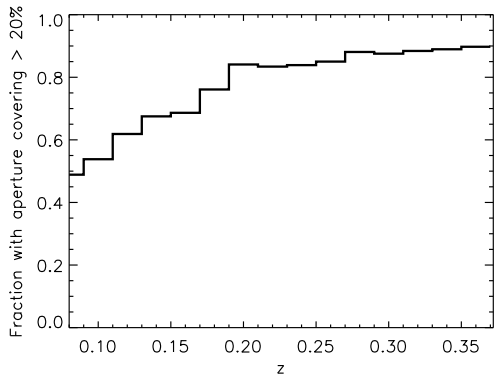}
 \caption{The fraction of galaxies where the 1\farcs5 aperture
 includes $>20\%$ of the galaxy light.}
 \label{cover}
 \end{center}
\end{figure}

We compute the covering
fraction from the difference between the 1\farcs5 aperture magnitude and the
total galaxy magnitude.  Most galaxies in the sample have a covering fraction
$>20\%$; Figure~\ref{cover} shows the fraction of galaxies with
covering fraction  $>20\%$ as a function of redshift for the galaxies
(both paired and unpaired) at $z=0.080-0.376$.

\citet{fabricant08} conduct a detailed investigation of aperture effects
in their sample of overlapping Hectospec and SDSS spectra,  a 
subset of our full spectroscopic survey.  SDSS spectra are acquired through a
3\arcsec\ aperture. Thus \citeauthor{fabricant08} scale the  Hectospec line fluxes
by the 3\arcsec\ aperture DLS $R$-band magnitude to make direct comparisons
with the SDSS line fluxes.  They conclude that the median [O II] 3727~\AA~ and
 H$\alpha$ line fluxes from Hectospec agree to within $3-4\%$ of the SDSS
values, and that the scatter in the line fluxes is dominated by systematic
uncertainties.  The $3-4\%$ difference between Hectospec and SDSS line
flux measurements is small compared to the $23\%$ RMS scattter in the 
H$\alpha$ line flux from Hectospec.

The agreement between
the Hectospec and SDSS data implies that the H$\alpha$ star formation rate
within apertures of 1.5 and 3\arcsec\ scale with the $R$-band galaxy luminosity.
The relative agreement between the scaled Hectospec and SDSS H$\alpha$
line flux is independent of the absolute magnitude 
(see Figure 13 in \citealp{fabricant08}), 
apparent magnitude, redshift, and $25\%$ light radius of the galaxies.
Westra et al.\ (2009, in prep.) similarly find that the H$\alpha$ 
luminosity determined from the 1\farcs5 fiber and $R$-band 
magnitude is consistent with the H$\alpha$ luminosity from the 
3\arcsec\ aperture photometry and $R$-band magnitude of \citet{shioya08}.

Like the H$\alpha$ luminosity, we assume that $D_n4000$ observed in the 1\farcs5 
aperture represents the galaxy as a whole.  \citet{fabricant08} show that the Hectospec 
and SDSS measurement of $D_n4000$ agree very well over the scales observed (1.4 to 
2.8~kpc radius at $z=0.1$).  This agreement again suggests that measurement of 
$D_n4000$ for the stellar population included in the Hectospec aperture is not 
significantly affected by systematic biases compared to the population 
observed in the larger SDSS aperture.

\section{Pair and Field Sample Selection}  \label{sample}

We measure the effects of galaxy interactions on the
star formation activity and other galaxy properties at intermediate redshift.
We use two pair samples derived from the 6,935 galaxies  with spectroscopy and 
$R<20.3$ in the redshift range $z=0.080-0.376$.  

The first sample, the ``full'' major pairs sample, includes
 all galaxies in major pairs that meet our projected  spatial and line-of sight 
peculiar velocity criteria (\S\ref{fullmaj}) in the range $z=0.080-0.376$. There 
are 622 galaxies in the full major pairs sample.

 The second sample, the ``volume-limited'' major 
pairs sample, is a subset of the full major pairs sample, and is restricted to
galaxies that meet both redshift and absolute luminosity selection criteria 
(\S\ref{majpair}).  The volume-limited major pairs sample allows us to study
the strength, frequency, and timescale of triggered star formation.
The volume-limited major pairs sample is well suited to comparison with the 
predictions of the numerical simulations of \citet{dimatteo08}, who measure the 
intensity, frequency, and  duration of merger-driven star formation in their large 
suites of numerical simulations of major interactions.  The volume-limited major
pairs sample contains 339 galaxies.

Our pair samples are the largest spectroscopic samples of pairs to date at
intermediate redshift ($z\sim 0.1 - 0.4$).  In comparison, the Millennium Galaxy Catalog 
\citep[MGC;][]{liske03,driver05,allen06} is a $96\%$ complete spectroscopic survey of 
10,095 galaxies to $B_{MGC} < 20$ mag.  The MGC is similar in number and completeness 
to our sample but is at lower redshift.  \citet{depropris07} study pair galaxies 
in the MGC: they identify a volume-limited sample 
of 3,237 galaxies in the range  $0.010<z<0.123$, including 112 galaxies in
 pairs with projected separation $\Delta D < 28$~kpc [20~$h_{100}^{-1}$kpc].
The CNOC2 Redshift Survey contains redshifts for $\sim5,000$ galaxies at
$0.1<z<0.6$ with a cumulative completeness of $50\%$ and differential 
completeness of $20\%$ at the limit $R_c \leq 21.5$ \citep{yee00,patton02}.
\citet{patton05} study properties of dynamically close pairs in the CNOC2
survey.  Our sample covers the same redshift range  but is substantially
 more complete than CNOC2.  Table~\ref{tab_samples} lists the selection criteria and 
the number of galaxies  in the  full major pair sample and in the volume-limited 
major pair sample.

\begin{deluxetable}{lll}
\tablecolumns{3} 
\tablewidth{0pc} 
\tablecaption{Criteria and size of galaxy pair samples.
 \label{tab_samples} }
\tablehead{
\colhead{Criteria}   & \colhead{Full sample} &  \colhead{Volume-limited sample} }
\startdata
\colhead{Major pairs\tablenotemark{a}}                       & 622 & 339  \\
\colhead{Non-AGN\tablenotemark{b}}                           & 601 & 324   \\
\colhead{$D_n4000 < 1.44$\tablenotemark{c}}                  & 302 & 134  \\
\colhead{$N_c \leq 4$\tablenotemark{d}}                      & 214 & 72   \\
\enddata
\tablenotetext{~}{Note.--  The sample size includes galaxies that meet listed condition 
and all previously listed conditions.}
\tablenotetext{a}{$\left | \Delta M_R \right |<1.75$.}
\tablenotetext{b}{AGN classification in \S\ref{agn}.}
\tablenotetext{c}{$D_n4000$ classification in \S\ref{d4000}.}
\tablenotetext{d}{~Local density measure $N_c$ described in \S\ref{density}.}
\end{deluxetable}

\subsection{Full major pairs sample} \label{fullmaj}

We select galaxy pairs with projected spatial separation $\Delta D \leq 70$~kpc, 
corresponding to the limit $\Delta D \leq 50$~h$_{100}^{-1}$kpc  commonly applied 
in previous studies  \citep[e.g.][]{bgk}.  We require a line-of-sight peculiar 
velocity difference
$\Delta V / (1 + z) < 500$~km~s$^{-1}$.  The limit of $\Delta D \leq 70$~kpc 
includes pair galaxies with the most significantly enhanced EW(H$\alpha$) 
\citep[e.g][]{bgk}, while minimizing the presence of interlopers.  The $\Delta V$ 
limit is motivated by  \citet{bgk}, who find
no significantly enhanced EW(H$\alpha$) emission in  pairs with 
$\Delta V \gtrsim 500$~km~s$^{-1}$ (their Fig. 2b) in their sample of 786 galaxies 
from the CfA2 Redshift Survey, and is consistent with previous pair studies 
\citep{depropris07,woods06}.  

We exclude potential pair galaxies from our sample if either member of the pair lies 
within a projected $70$~kpc from the survey boundary or the edge of the region 
excluded around a bright star.  For galaxies in a compact group, we compare each galaxy 
to its nearest neighbor; odd numbers of ``pair'' galaxies can occur in compact groups.

We compute the total pair fraction for all pairs -- major and minor -- in the
redshift interval $z=0.080-0.376$.   There are 809 galaxies in major or minor pairs
 or compact groups that meet the projected spatial and  peculiar velocity requirements.
  The total pair fraction is $12\%$ (809/6,935).

\begin{figure}[htb]
  \begin{center}
 \includegraphics[width=2.25in,angle=90]{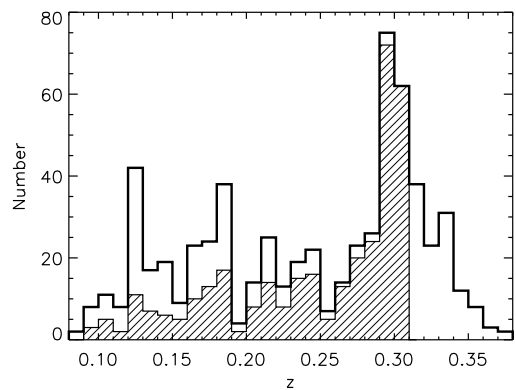}
 \caption{Redshift distribution for the full major pair galaxies (open)
 and for the volume-limited major pair galaxies (filled).}
 \label{full_zdist}
 \end{center}
\end{figure}

We restrict our sample to major pairs with $\left | \Delta M_R \right | < 1.75$,
corresponding to a luminosity (approximate mass) ratio $>1/5$.  Simulations of interacting
galaxies predict that major interactions are more effective than minor 
interactions at triggering star formation; only the fainter companion in a
minor interaction occasionally shows triggered star formation \citep{cox08}.  
Observations of major and minor pair galaxies support this predicted behavior
(\citealp{dasyra,woods07,ellison08};  Westra et al. 2009, in preparation.)   
The full major pair sample includes 622 galaxies in major pairs in the range 
$z=0.080-0.376$.  Figure~\ref{full_zdist} shows the 
redshift distribution and Figure~\ref{hist_sepr1} shows the distribution of 
projected separation for the full major pairs sample. Table~\ref{full_tab}
lists the full major pair sample galaxies and their properties.

\tiny
\begin{deluxetable}{llllllllll}
\tablecolumns{10} 
\tablewidth{0pc} 
\tablecaption{Sample of major galaxy pairs\tablenotemark{a,b}
 \label{full_tab} }
\tablehead{ 
\colhead{Pair} & \colhead{RA\tablenotemark{c}} & \colhead{Dec\tablenotemark{c}} & \colhead{$z$\tablenotemark{d}} 
 & \colhead{$m_R$\tablenotemark{d}} & \colhead{$M_R$\tablenotemark{d,e}} & \colhead{$\Delta D$} & 
\colhead{$\Delta V$\tablenotemark{d}}  & \colhead{$D_n4000$\tablenotemark{d}} & 
\colhead{SSFR$_{H\alpha}$\tablenotemark{d}} \\
 & & & & &      & \colhead{(kpc)} & \colhead{(km~s$^{-1}$)} & & 
 \colhead{(M$_{\sun}$~yr$^{-1} / 10^{10}$~M$_{\sun}$)}}
\startdata 
001 & 9:14:52.622 & 30:05:40.858 & 0.26472 & 19.56 & -21.23 &  68 &  388 &  1.29 &  1.36\\
001 & 9:14:53.920 & 30:05:40.377 & 0.26309 & 19.97 & -20.76 &  68 &  388 &  1.79 &  0.10\\
\tableline
002 & 9:14:52.849 & 30:17:05.567 & 0.12805 & 20.10 & -18.80 &  50 &   51 &  0.96 &  1.26\\
002 & 9:14:53.528 & 30:17:25.933 & 0.12813 & 19.17 & -19.74 &  50 &   51 &  1.05 &  3.99\\
\tableline
003 & 9:14:53.583 & 30:31:44.906 & 0.25643 & 20.20 & -20.40 &  65 &  108 &  1.14 &  1.18\\
003 & 9:14:54.188 & 30:31:30.369 & 0.25688 & 19.60 & -21.14 &  65 &  108 &  0.00 &  0.04\\
\tableline
004 & 9:14:57.683 & 29:57:15.864 & 0.24721 & 19.08 & -21.44 &  26 &  133 &  1.07 &  3.83\\
004 & 9:14:58.198 & 29:57:15.988 & 0.24780 & 19.21 & -21.34 &  26 &  133 &  1.11 &  3.79\\
\tableline
005 & 9:14:57.762 & 29:49:34.947 & 0.18074 & 18.58 & -21.20 &  26 &  171 &  1.44 &  0.36\\
005 & 9:14:58.434 & 29:49:34.583 & 0.18014 & 17.81 & -22.01 &  26 &  171 &  1.73 &  0.00
\enddata
\tablenotetext{a}{Full table available online: http://tdc-www.cfa.harvard.edu/instruments/hectospec/progs/tsf/.}
\tablenotetext{b}{Standard $\Lambda$CDM cosmology:
$H_0 = 71$~km~s$^{-1}$~Mpc$^{-1}$, $\Omega_m = 0.3$, and $\Omega_{\Lambda} = 0.7$.}
\tablenotetext{c}{J2000 coordinates.}
\tablenotetext{d}{Typical error estimates: \\
$cz: 34$~km~s$^{-1}$ emission line galaxies, 65~km~s$^{-1}$ absorption line galaxies (\S\ref{specmeas}),\\
$m_R: 0.05$ mag (\S\ref{phot}),\\
 $D_n4000: 4.5\%$. (\S\ref{specmeas}), \\
$M_R: \sim0.1$~mag, varies with redshift and galaxy type (\S\ref{phot}), \\
SSFR$_{H\alpha}: 30\%$ (\S\ref{sfr}).}
\tablenotetext{e}{k+e corrected.}
\end{deluxetable}

\normalsize

Visible evidence of tidal interactions in the DLS combined $BVR$-band images, such 
as tidal tails, marked asymmetries, or extended halos, can be seen in  
$\sim 1/4$ of the major pair galaxies. 
Analyzing the morphological features in detail is beyond the scope of this 
project.  The images are available on our website$^1$.

\subsection{Volume-limited major pairs sample} \label{majpair}

We construct a volume-limited sample of major galaxy pairs to 
study trends in triggered star formation and other galaxy properties.  
The volume-limited sample contains galaxies in the redshift 
range $0.080<z<0.310$ and with magnitude $M_R < -20.8$.   The  limits
 on $M_R$ and $z$ maximize the number of  galaxy pairs in the
volume-limited sample.  The lower redshift limit, $z=0.080$, minimizes 
aperture effects (\S\ref{aperture}).
There are 339 galaxies in major pairs in the volume-limited sample.

\begin{figure}[htb]
  \begin{center}
 \includegraphics[width=2.25in,angle=90]{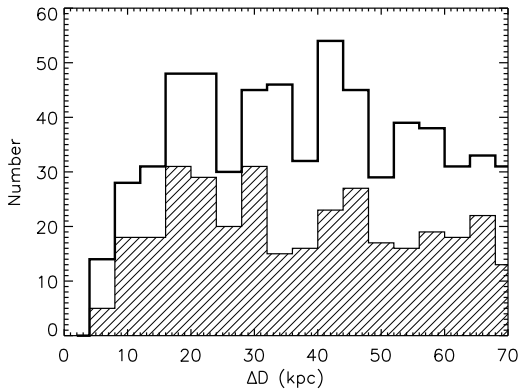}
 \caption{Distributions of projected separation for galaxies in the full
major pair sample (open) and for the volume-limited major pair sample (filled).   
The distribution is nearly flat from $10<\Delta D <70$~kpc, as expected.  
At very small $\Delta D$, not all pairs are resolved. }
 \label{hist_sepr1}
 \end{center}
\end{figure}

\begin{figure}[htb]
  \begin{center}
 \includegraphics[width=2.25in,angle=90]{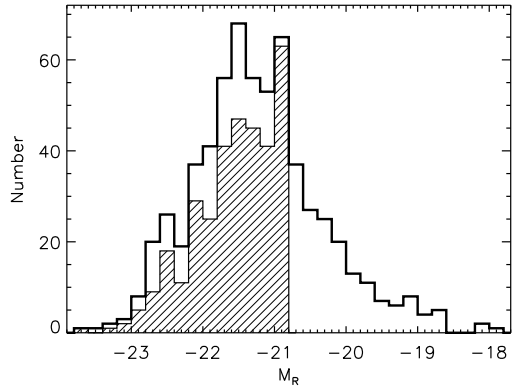}
 \caption{Absolute magnitude distribution for galaxies in the full major
pair sample (open) and the volume-limited major pair sample 
(filled). \label{hist_mag1} }
 \end{center}
\end{figure}

The distribution of projected separations (Figure~\ref{hist_sepr1})
is relatively flat from $15<\Delta D <70$~kpc, as expected from
the galaxy correlation function \citep[e.g.][]{estrada}.  At separations 
$< 15$~kpc, galaxies in the DLS images may not be resolved into separate 
systems.  Figures~\ref{hist_mag1} shows the
distribution of absolute magnitudes of galaxies in the full major
pair sample and in the volume-limited major pair sample.

\subsubsection{Volume-limited unpaired field sample} \label{fieldsample}

We construct a volume-limited sample of galaxies with no detected companions 
(the ``field'' sample) for comparison with the volume-limited pair sample.
 The field sample includes only galaxies without companions that
satisfy the pair selection criteria.
 The field galaxies in the volume-limited sample meet the same
magnitude ($M_R<-20.8$) and redshift ($z=0.080-0.310$) requirements as the
major pair galaxies.   The range of magnitude differences excluded 
from the field galaxies ($\left | \Delta M_R \right | < 2$) is slightly broader 
than the range accepted for major pairs ($\left | \Delta M_R \right | < 1.75$)
to exclude systems that could marginally be considered major pairs.
We exclude galaxies within $70$~kpc from the survey boundary or 
$70$~kpc from the boundaries around bright stars.
The volume-limited field sample includes 2,234 galaxies.

\begin{figure}[htb]
  \begin{center}
 \includegraphics[width=2.25in,angle=90]{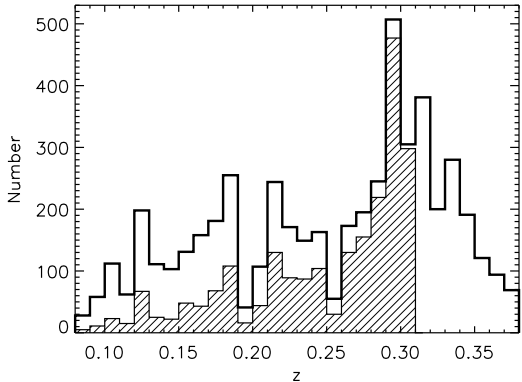}
 \caption{Redshift distribution for the field galaxies (open) and the 
volume-limited field galaxies (filled). \label{zhist_fld} }
 \end{center}
\end{figure}

The distributions of $M_R$ in the volume-limited major pair sample and 
 volume-limited field sample are similar.  We use the Kolmogorov-Smirnov 
(K-S) test to determine that the two distributions are drawn from the sample parent
sample ($P_{KS} = 0.37$).  The volume-limited major pair sample
 and the unpaired field sample also have similar distributions of redshifts 
($P_{KS} = 0.29$).  Figure~\ref{zhist_fld} shows the redshift distribution
 of field galaxies.

Some of the field galaxies may have a major companion fainter than the survey 
limit.  We estimate the fraction of field galaxies likely to have an undetected major 
companion.  The fraction of galaxies in pairs in the volume-limited sample
and with apparent magnitude $m_R < 18.55$ is $11\%$ (102/902).  
The $m_R <18.55$ limit allows  identification of
pair galaxies with companions fainter by up to 1.75 mag.  There are 1633 out of 
2220 field galaxies ($74\%$) in the range $18.55<m_R<20.3$ that could have a major 
companion fainter than our survey limit. An additional factor of 0.5 accounts for a 
pair galaxy having a $50\%$ probability of being brighter than its 
companion.  Hence the fraction of field galaxies  likely to have an undetected
major companion is $\sim4\%$. 
There are also some field galaxies in later stages of merging such that 
a companion is unresolved.  Because the field sample contains some pairs
 and merger remnants, any differences in galaxy properties between the pairs 
and field samples attributable to  interactions are lower limits.

\section{Spectroscopic Identifiers} \label{properties}

In this section we describe galaxy classification metrics 
to identify candidates for tidally triggered star formation.
We identify the galaxies with strong AGN to remove them from our sample of
star-forming galaxies (\S\ref{agn}).  We describe the use of the spectroscopic
indicator $D_n4000$ as a galaxy classification tool (\S\ref{d4000}).

\subsection{Identification of AGN}  \label{agn}

We exclude strong AGN from our analysis because measurements of
emission lines, such as H$\alpha$, would not accurately represent star 
formation activity in galaxies with active nuclei.
We distinguish narrow-line AGN from star forming galaxies
with measurements of the emission lines H$\alpha$, H$\beta$, 
[N II], and [O III] using the classification
metric of \citet{kewley06},  based on the ``BPT'' diagrams of \citet{bpt}.
The line fluxes H$\alpha$ and H$\beta$ are stellar absorption corrected.

\begin{figure}[htb]
  \begin{center}
 \includegraphics[width=2.25in,angle=90]{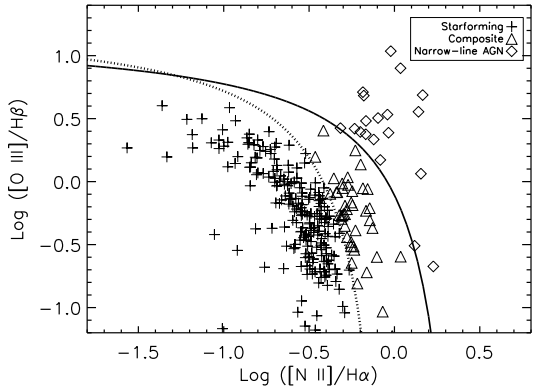}
 \caption{Classification of pair galaxies as  starforming, composite, or narrow-line
 AGN using
 the \citet{kewley06} criteria.  The solid black line shows the boundary between
composite galaxies and AGN, and the dotted line shows the boundary between
star forming and composite galaxies.  We include composite galaxies in our sample
for analysis of star formation activity.}
 \label{classify_p}
 \end{center}
\end{figure}

  The requirements we place on the spectra for
classification are very mild: EW(H$\alpha) > 2$~\AA~and EW(H$\beta) > 2$~\AA.
We impose no requirement on [N II] or [O III] because galaxies without 
measurable [N II] or [O III] could only have a very weak AGN. Furthermore, 
 galaxies with EW(H$\alpha) < 2$~\AA~have minimal star formation.
Figure~\ref{classify_p} shows the classification diagram
for the pair galaxies.

We include composite galaxies in our pair and field samples, 
but exclude the galaxies classified as AGN.  
Although these criteria may  admit weak AGNs, most of the 
contribution to the H$\alpha$ emission should be from star-forming regions.
There are 119 narrow-line AGNs in the $z=0.080-0.376$ sample, 19 of which 
are in major pairs (\S\ref{majpair}).

 Our ability to detect narrow-line AGN within the host galaxy declines
with redshift because the covering fraction of the spectroscopic fiber increases
with redshift.  At redshift $z=0.080$, the 1\farcs5 fiber diameter corresponds to
2.2~kpc, and at $z=0.310$, the 1\farcs5 fiber covers 6.8~kpc.
We also admit fewer AGN at higher redshift into our spectroscopic sample because
objects with stellar light profiles are excluded from our spectroscopic catalog.

Despite these caveats to our ability to detect narrow-line AGN at higher 
redshift, the mean redshift of the narrow-line AGN in the full spectroscopic 
sample (both pair and unpaired galaxies) is higher than that of the objects 
classified as starforming or composite: mean $z=0.27$ and $z=0.23$, 
respectively.  The redshift distributions differ 
significantly ($P_{KS}=1.6\times10^{-5}$).  We attribute this trend to the 
evolutionary decline in AGN activity at low redshift \citep[e.g.][]{brand,cavaliere}.

We identify galaxies with broad emission lines by visually inspecting all 
spectra that are potential broadened line objects.
The criteria for broad-line AGN candidates is EW(H$\alpha) > 3$~\AA~ and 
the line flux in a band widened from $\pm 8$~\AA~ to $\pm 12$~\AA~ increases 
the measured H$\alpha$ flux by more than $10\%$, {\it or}  
EW([N II]) $> 3$~\AA~ and  the line flux in a band widened  from $\pm 8$~\AA~ to 
$\pm 12$~\AA~ increases the measured [N II]  flux by more than $10\%$.  We 
identify 26 galaxies with broadened lines in the $z=0.080-0.376$ sample.  Five of 
these are in major pairs.  

 We exclude galaxies classified as AGN from our analysis of star formation 
activity; we retain the non-AGN companion.  
Table~\ref{agn_table} reports the AGN fraction in each pair sample  and in the matching 
unpaired field sample.  The AGN fraction in pairs exceeds that in the matching field sample for
each of the subsamples.  The increase in AGN fraction for both the full and 
volume-limited major pair samples is a factor of $\sim2$, a significance of $\sim2\sigma$.
The higher AGN fractions in the volume-limited samples (pair and field) compared to the
 full samples is consistent with the exclusion of lower luminosity galaxies from
the volume-limited sample, which are less likely to contain an AGN.

\begin{deluxetable}{lllll}
\tablecolumns{5} 
\tablewidth{0pc} 
\tablecaption{AGN fraction
 \label{agn_table} }
\tablehead{
\colhead{Sample} & \colhead{Pair\tablenotemark{a}} &  \colhead{Field\tablenotemark{b}} & \colhead{Significance} & \colhead{Pair sample size}}
\startdata
\colhead{Full}     & $8  \pm 2\%$   & $4 \pm 2\%$ & $2\sigma$ & 280  \\
\colhead{Volume limited} & $12 \pm 3\%$   & $5 \pm 2\%$  & $2\sigma$ & 128  \\
\enddata
\tablenotetext{a}{Fraction of AGNs compared to fraction of galaxies that 
meet our criteria for classification (\S\ref{agn}). We identify only very 
bright AGNs with our spectra and our classification criteria.}
\tablenotetext{b}{Proportional representation of field galaxies matching the
 $M_R$ and redshift distribution of the pair sample.}
\end{deluxetable}

\subsection{Galaxy classification by $D_n4000$} \label{d4000}

It is useful to segregate galaxies into two categories: early-type
galaxies, which are generally gas-poor and have little or no active
star formation, and late-type, which contain more gas and have young
stellar populations. The gas-rich systems are potentially susceptible to 
tidally triggered star formation in major interactions
\citep[e.g.][]{mihos+hern96,tissera,cox06,dimatteo07}.
 Interactions between gas-poor galaxies (``dry mergers'') produce little or no star
 formation activity, although they contribute substantially to the build-up of 
massive galaxies  \citep[e.g.][]{tran05,vandokkum05,cattaneo08}.
Our analysis of star formation activity in pair galaxies focuses on the
late-type systems.

 There are a number of classification schemes to separate the
early and late-type galaxy populations.  Photometric discriminants include color, 
concentration, and absolute magnitude \citep[e.g.][]{strateva01,kauff03_54},
and spectroscopic methods include $D_n4000$ and H$\delta$ absorption
\citep{kauff03_33}.  The $D_n4000$ indicator discriminates by  stellar population age.
    At wavelengths bluer
than 4000~\AA, metal lines in low mass stars absorb the light and cause a 
``break'' in the spectrum.   As the stellar population ages and the massive, 
hot stars die off, $D_n4000$ increases monotonically with time.
\citet{kauff03_33} use stellar population models to show that galaxies with
$D_n4000 \lesssim 1.5$ have young stellar populations ($\lesssim 1$~Gyr).
Metallicity has a strong effect on the the value of $D_n4000$ only after 
1 Gyr  past a burst of star formation \citep[see Fig. 2 in][]{kauff03_33}. 
Measurement of $D_n4000$ is insensitive to galaxy reddening.

\citet{vergani08} use $D_n4000$ to separate
spectroscopic early-type galaxies from late-type galaxies at a dividing line of
$D_n4000 = 1.5$ in their analysis of galaxy stellar mass assembly in the VIMOS
VLT Deep Survey. \citet{mignoli05} also use  $D4000=1.6$ as a dividing line 
(different definition of $D4000$ from \citealp{bruzual83}: ratio of 
flux in bands 4050-4250~\AA\ to 3750-3950~\AA) along with other spectral 
measurements to  classify galaxies in the K20 survey, 
a near-IR selected redshift survey.   We follow this approach.

\begin{figure}[htb]
  \begin{center}
 \includegraphics[width=2.25in,angle=90]{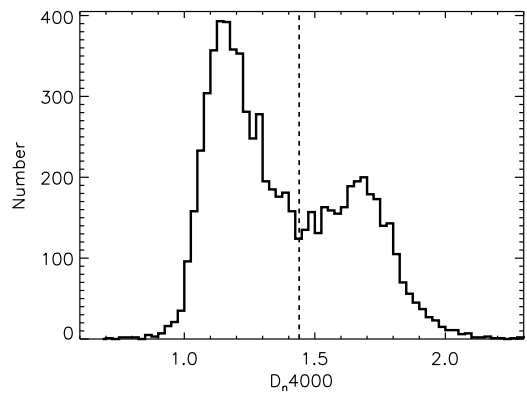}
 \caption{Distribution of $D_n4000$ for all galaxies in the spectroscopic sample.  
The bimodal distribution has a local minimum at $D_n4000 = 1.44$ (dashed line).}
 \label{hist_d4000}
 \end{center}
\end{figure}

The distribution of  $D_n4000$ is bimodal, separating galaxy populations
dominated by old stars from systems with recent star formation.
Figure~\ref{hist_d4000} shows the distribution of $D_n4000$ for the
6,644 galaxies (both pair and unpaired) at $z=0.080-0.376$ with a robust
$D_n4000$ measurement.  The peaks are at $\sim 1.15$ and $\sim 1.75$.
\citet{kauff03_33} similarly observe a bimodal distribution in $D_n4000$, 
with peaks at 1.30 and 1.85 in their sample of $\sim100,000$ galaxies in the SDSS.
We choose the minimum between  the bimodal distribution 
 as our dividing line, $D_n4000 = 1.44$ (Figure~\ref{hist_d4000}).
We refer to galaxies with $D_n4000 \leq 1.44$ as ``low'' $D_n4000$ galaxies and 
those with $D_n4000>1.44$ as ``high'' $D_n4000$ galaxies.
 Our analysis of star formation activity includes  the low $D_n4000$ 
galaxy in a mixed pair, but does not require both galaxies to have
$D_n4000 < 1.44$.

\begin{figure}[htb]
  \begin{center}
 \includegraphics[width=2.25in,angle=90]{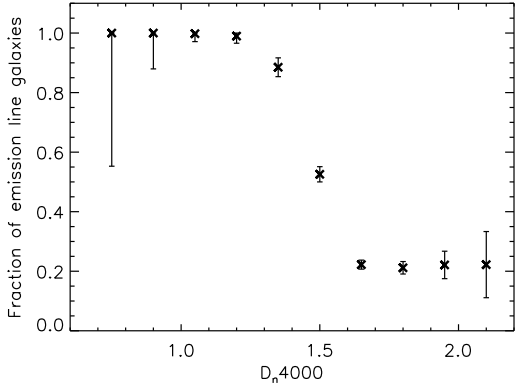}
 \caption{Fraction of emission-line galaxies  as a function of $D_n4000$
for all galaxies in the spectroscopic sample, excluding AGN.  Emission line 
galaxies have EW(H$\alpha) \geq 3$ or  EW([O II])$\geq 3$.
There is a steep decline in the fraction between $D_n4000 = 1.3-1.5$.
Error bars from bootstrap resampling indicate the $95\%$ confidence intervals.}
 \label{emfracd4}
 \end{center}
\end{figure}

The emission line fraction is a strong function of $D_n4000$.  
Figure~\ref{emfracd4} shows the steep decline in the fraction 
of  emission line galaxies between $D_n4000 = 1.3-1.5$.  The emission line 
galaxies have EW(H$\alpha) \geq 3$ or  
EW([O II])$\geq 3$.  Thus segregating by $D_n4000$ is reasonable and 
corresponds well to segregating by the presence of emission lines.

\begin{figure}
\begin{center}
 \includegraphics[width=2.25in,angle=90]{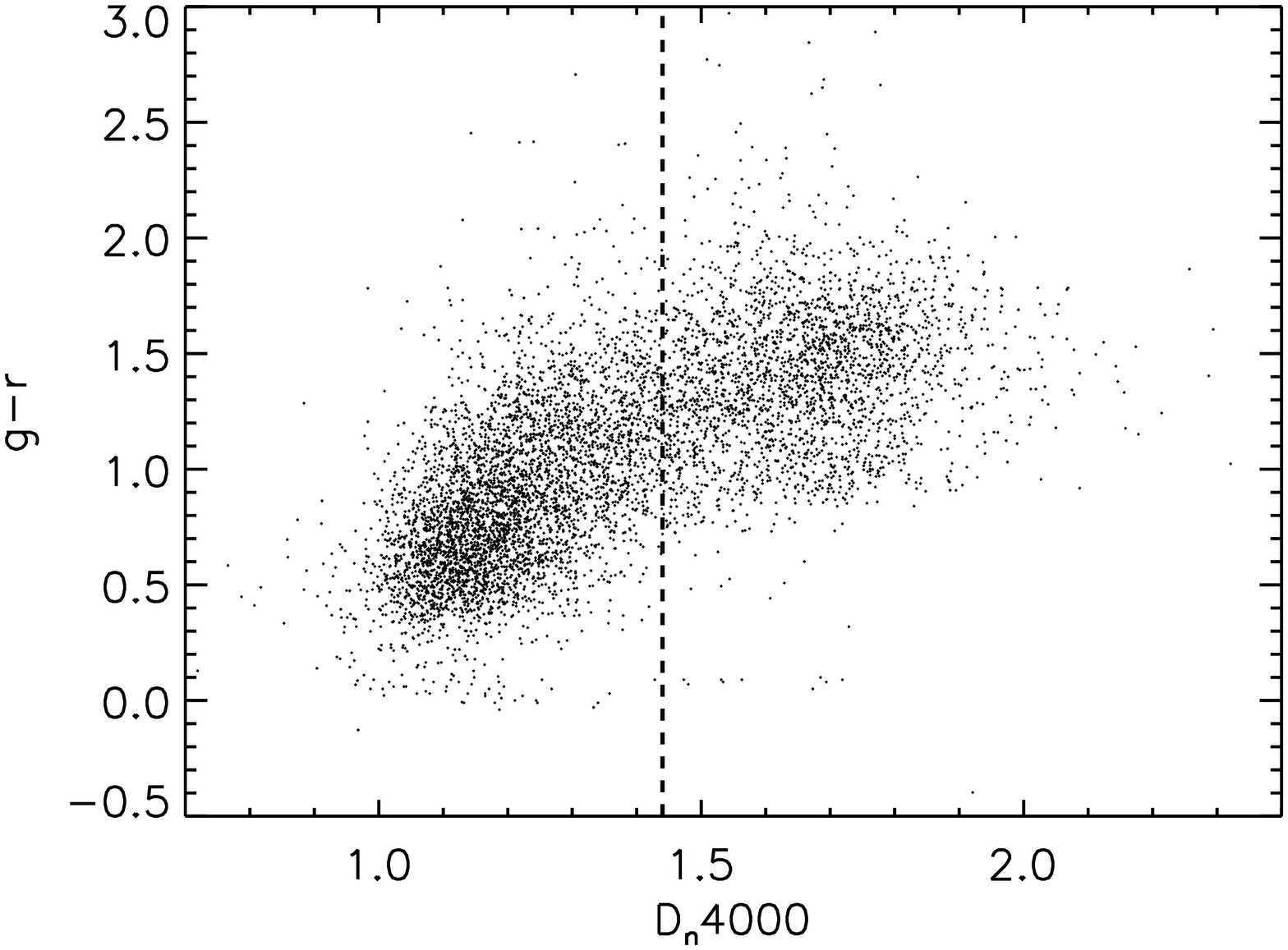}
 \caption{$D_n4000$ versus rest-frame ($g-r$) color.  
 Two galaxy populations are visible: systems with blue colors (lower left),
 and systems with red colors (upper right).  The dashed line indicates 
$D_n4000 = 1.44$ (see Figure~\ref{hist_d4000}).}
 \label{clr_d4000}
 \end{center}
\end{figure}

We also compare the use of $D_n4000$ for galaxy classification with color,
 another widely used indicator of galaxy type. Figure~\ref{clr_d4000} shows 
the bimodal distribution of rest-frame SDSS ($g-r$) color versus $D_n4000$. 
We compute the rest-frame  SDSS ($g-r$) color using the k+e correction determined
by \citet{annis01} from the Pegase code \citep{pegase}.
 The separation by $D_n4000$ is more sharply defined than that of rest-frame color.  
The rest-frame galaxy colors are affected by reddening, and depend 
on the noisy k+e corrections.  $D_n4000$  provides 
a well-defined method to segregate galaxies and has the advantage that 
no redshift dependent corrections are required.

\section{Local Density Effects} \label{density}

Because pair galaxies are more common in higher density regions 
\citep{barton07}, and because
cluster and group galaxies are more likely to be red with less active
 star formation than isolated galaxies \citep[e.g.][]{hubble31,cooper,gerke}, 
comparisons between pair and field populations without attention to 
density can strongly bias the interpretation of studies of close pairs.
\citet{barton07} emphasize that including high density regions suppresses
the detection of triggered star formation.

To compare the environment of pair and field galaxies, we compute the number of companions  
($N_c$) within a co-moving sphere of radius 985~kpc centered on the galaxy 
(radius equivalent to 700 $h_{100}^{-1}$kpc).
 The 985~kpc radius is within the typical virial radius of 
clusters ($\sim 1$~$h_{100}^{-1}$Mpc, \citealp{rines03}), and maximizes the
survey area included in our analysis. This density measurement is consistent  with
 that of \citet{barton07}.  We measure $N_c$ within the 
volume-limited sample, requiring that the galaxy reside within $z=0.080- 0.310$ 
and have magnitude $M_R < -20.8$.  We count all neighboring galaxies, not just 
major companions.  We exclude regions within 985~kpc of the survey edge.

\citet{barton07} predict that most observed pair galaxies reside in 
higher mass cluster or group-sized halos, whereas field galaxies are usually in 
isolated low mass halos.
We test this prediction with our complete volume-limited data set, using $N_c$
as a proxy for halo membership.  Note that our $N_c$ counts neighbors; the 
halo count $N$ in Barton et al. counts total halo occupation, i.e. $N = N_c +1$.

\begin{figure}[htb]
  \begin{center}
 \includegraphics[width=5.25in,angle=90]{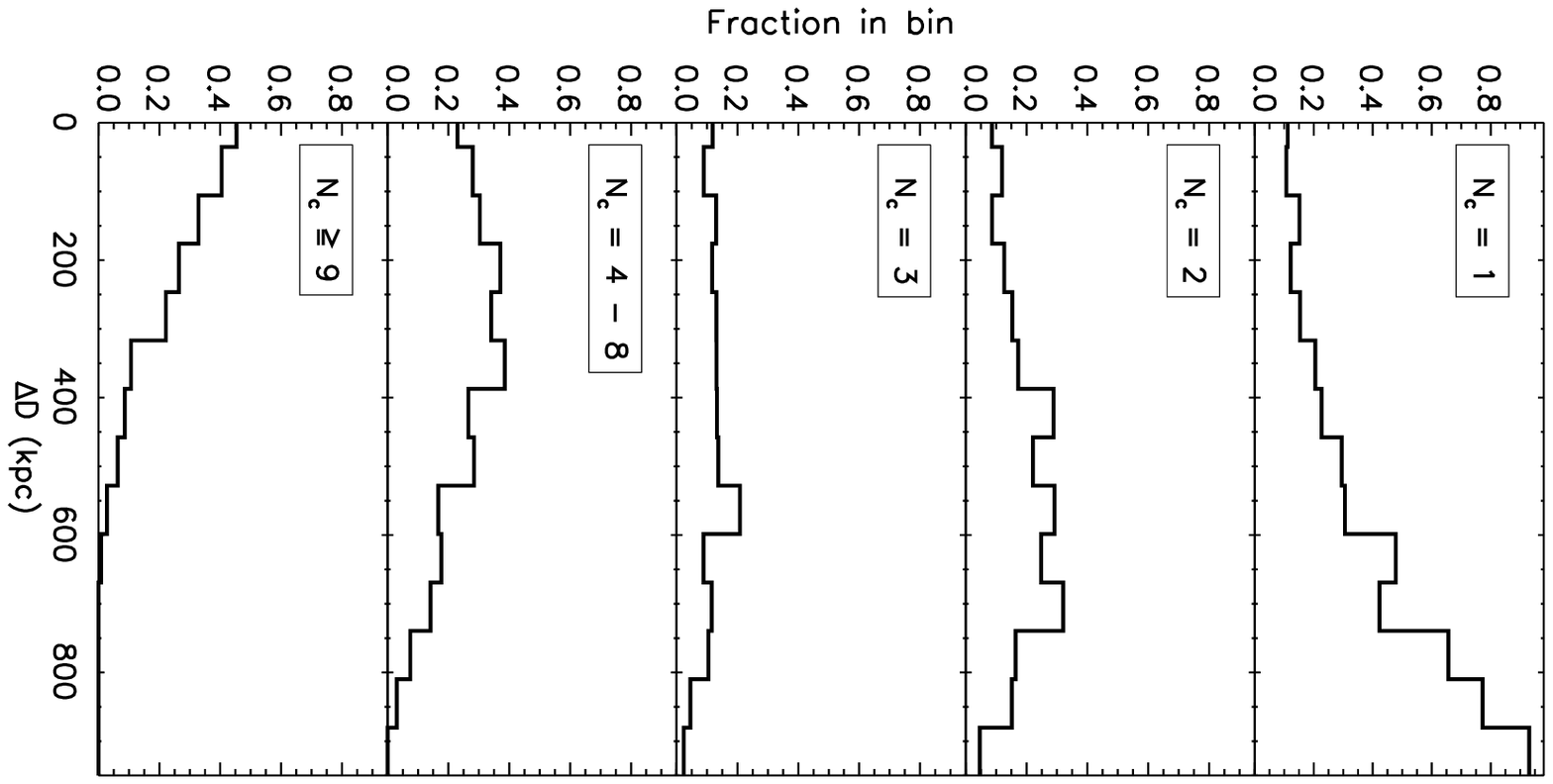}
 \caption{Fraction of galaxies in the volume-limited sample with a total 
$N_c$ neighboring galaxies within 985~kpc  co-moving (equivalent to 
700~$h_{100}^{-1}$kpc)  and $\Delta V < 1000$~km~s$^{-1}$ for the 
volume-limited sample  as a function of distance to  the nearest neighbor.  
Pair galaxies have $\Delta D < 70$~kpc.}
 \label{halos}
 \end{center}
\end{figure}

Figure~\ref{halos} shows the fraction of galaxies with various $N_c$
as a function of projected distance to the nearest neighbor.  Pair galaxies
fall in the range $\Delta D < 70$~kpc.  As \citeauthor{barton07} show,
pairs occur more frequently in locally dense
regions.  We find that  $32\%$ of pair galaxies have $N_c \geq 8$.
The overall fraction of galaxies in dense regions is smaller:
  $19\%$ of all galaxies lie in regions with $N_c \geq 8$.

Our observations are in excellent agreement with the 
 predictions of \citet{barton07}.  They find that $39\%$ of pair galaxies 
are in a host halo with total number of galaxies $N \geq 9$;  $19\%$ of
galaxies in the sample as a whole have $N \geq 9$ ($N_c \geq 8$).
Selecting a sample of field galaxies for a fair comparison of specific 
star formation rates   thus necessitates consideration of 
local density effects.

Selecting star forming galaxies with young stellar populations 
(low $D_n4000$) already reduces the effects of the environment because
young, blue galaxies are relatively more abundant in low density regions.  In the 
volume-limited major pair sample, $63\%$ of the low $D_n4000$ non-AGN 
galaxies reside in regions with $N_c \leq 4$.

Restricting the analysis of star formation activity solely by
$D_n4000$ does not, however, eliminate differences in star formation activity
with local density. 
The distribution of local densities differs between low $D_n4000$ pair and 
field galaxies ($P_{KS} = 1.6\times10^{-5}$).   
Thus we restrict any direct comparisons of specific star formation activity in
pairs and field galaxies to low density regions to ensure that density
effects do not dominate our measurements of star formation activity. A 
moderate limit of $N_c \leq 4$  maximizes our sample size.

\section{Measurements of tidally triggered star formation} \label{evidence}

Here we quantify the frequency, strength, and timescale of the triggered 
star formation for galaxies at intermediate redshift. We examine trends in star 
formation activity across redshift and luminosity for the galaxies in the
volume-limited sample.  The completeness and quality of spectrophotometry 
enable us  to carry out the most detailed spectroscopic analysis 
of pair galaxies to date in the redshift range $z=0.080-0.376$.
We examine the H$\alpha$ specific
star formation rate (SSFR$_{H\alpha}$, \S\ref{sfr}), the spectroscopic
parameter $D_n4000$ (\S\ref{d4000}), and a set of stellar population 
models (\S\ref{sfr}).

The volume-limited pair sample is well suited to comparison with
the predictions of numerical simulations.  We compare our
observations with the predictions of \citet{dimatteo08}, who  measure the 
intensity, frequency, and  duration of merger-driven star formation in their large 
suites of numerical simulations of major interactions.

\subsection{Star formation indicators and projected separation} \label{sfr_sep}

We examine  a set of star formation indicators as a function of projected separation 
to look for the signature of  triggered star formation at redshifts $0.080-0.376$.
  The star formation - $\Delta D$ 
anti-correlation frequently observed at low redshift 
\citep{bgk,lambas03,nikolic04,woods06,geller06,woods07,li08,ellison08} results 
from an increase in central star formation activity triggered by a close pass from a 
neighboring galaxy. As the pair galaxies move apart and the burst ages, the star 
formation activity decreases.

\begin{figure*}[htb]
  \begin{center}
 \includegraphics[width=2.25in,angle=90]{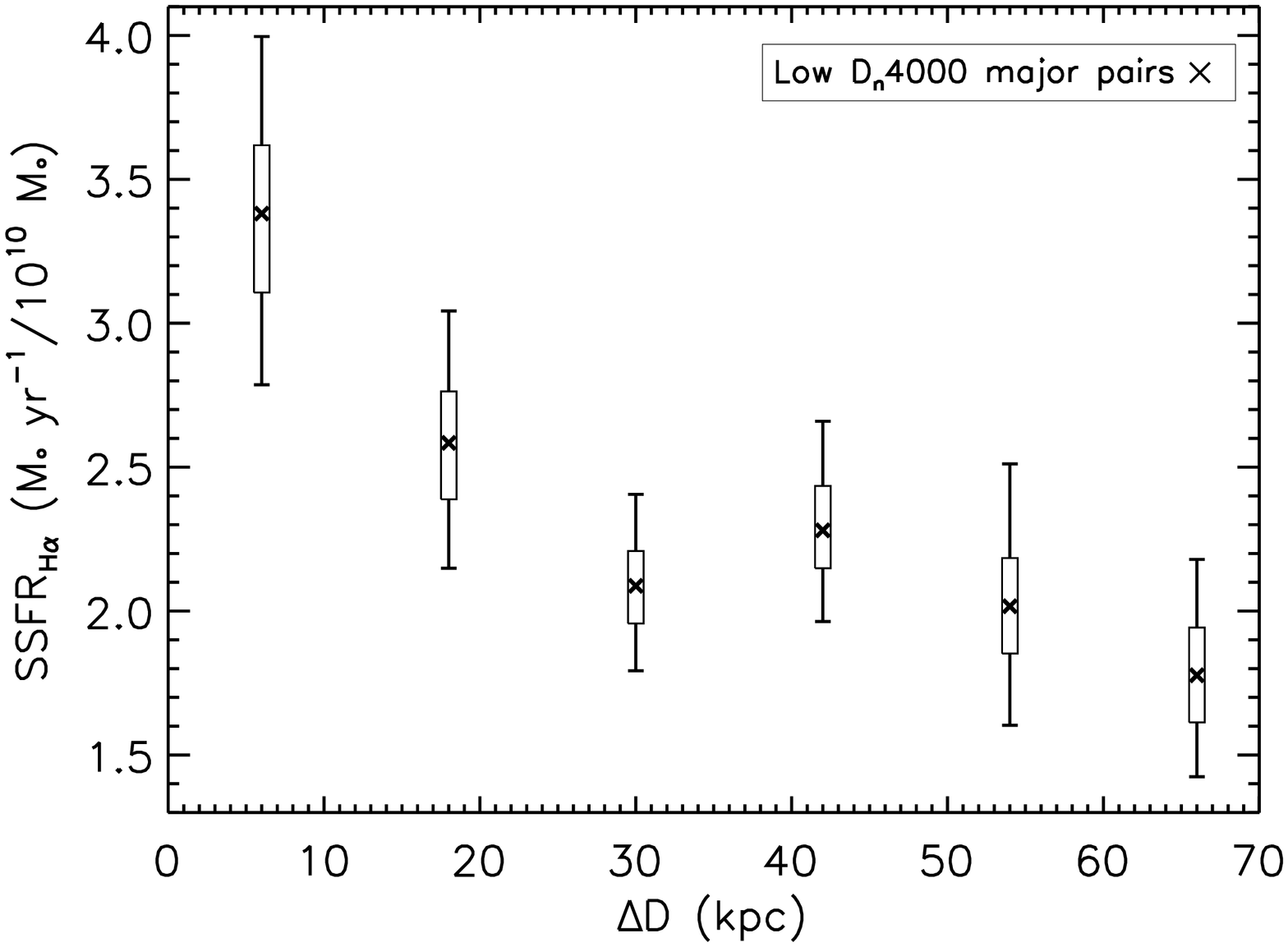}
 \includegraphics[width=2.25in,angle=90]{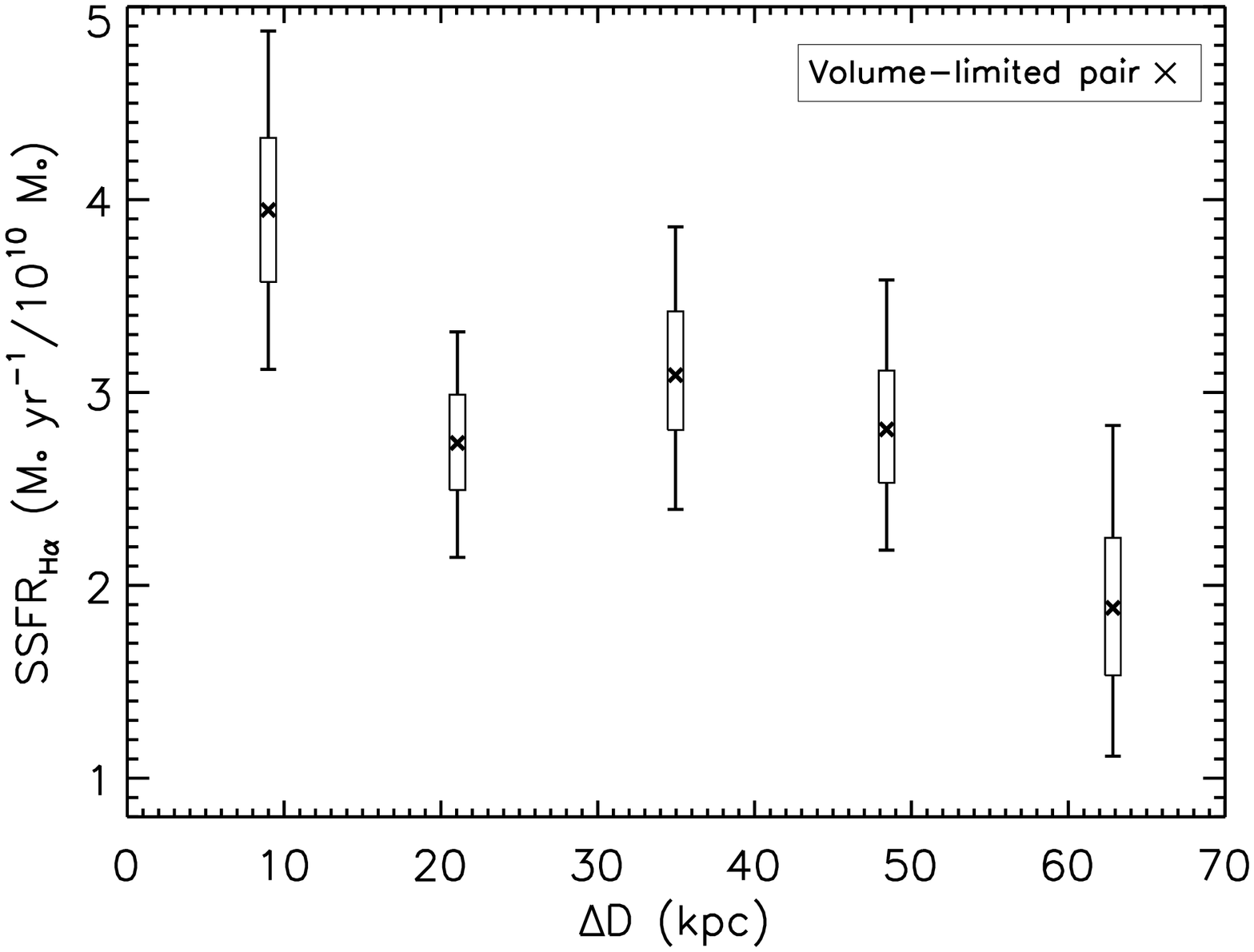}
 \includegraphics[width=2.25in,angle=90]{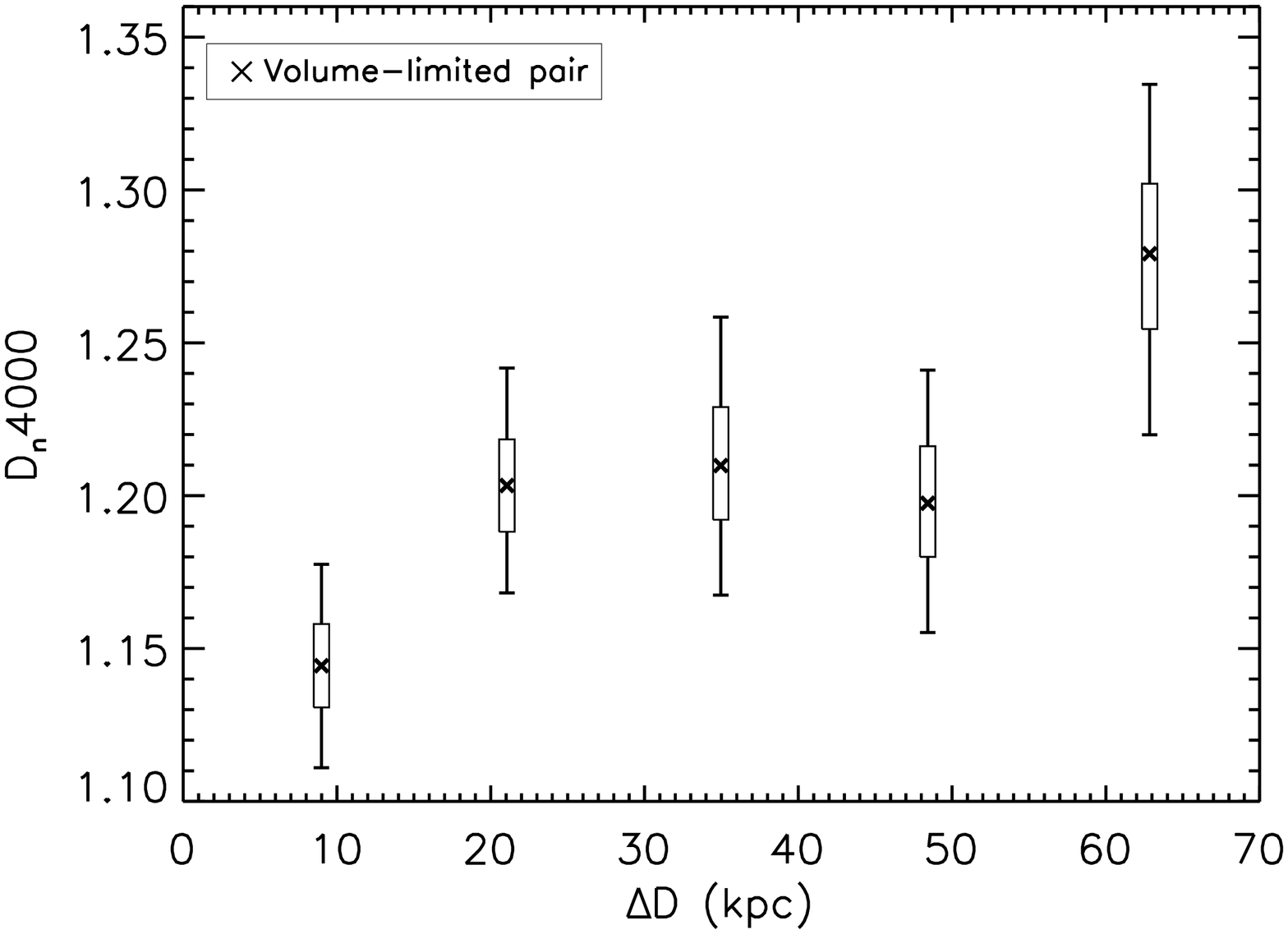}
 \includegraphics[width=2.25in,angle=90]{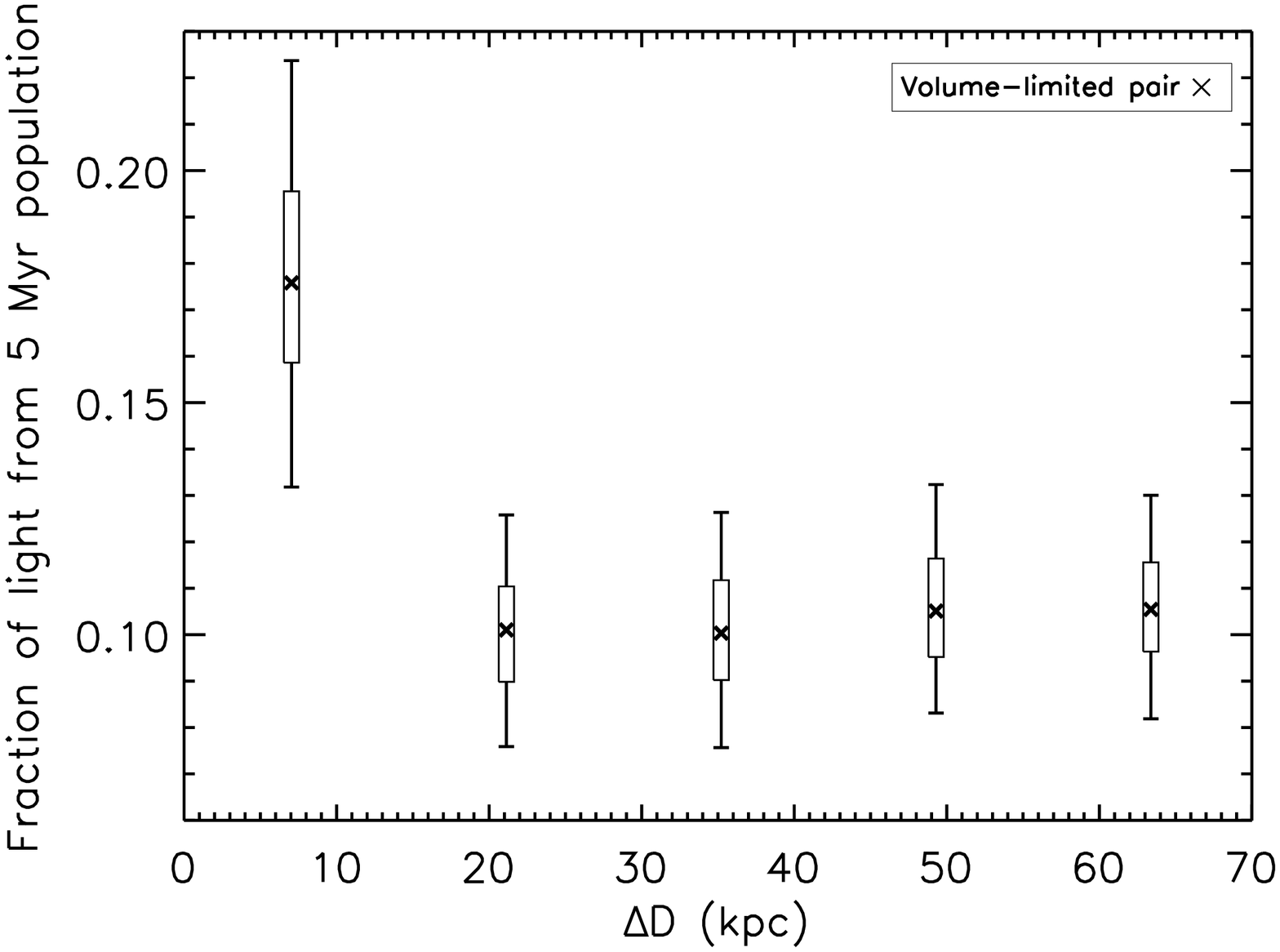}
 \caption{Mean star formation indicators versus projected separation.
We show the SSFR$_{H\alpha}$ versus projected separation for low $D_n4000$ major pair 
galaxies in the full sample at all local densities ({\it{top left}}).  Star formation 
indicators for the low $D_n4000$ galaxies in regions of low local density in the volume-limited
major pair sample: SSFR$_{H\alpha}$ ({\it{top right}}),  $D_n4000$ ({\it{bottom left}}), and 
fraction of light at 5500~\AA~from a 5~Myr starburst population ({\it{bottom right}}).
Error bars are from bootstrap re-sampling: boxes indicate the inter-quartile
range and the outer lines show $95\%$ confidence intervals.  For each star 
formation indicator shown above, the galaxies exhibit a correlation 
between the star formation indicator and $\Delta D$.}
 \label{hasepr1}
 \end{center}
\end{figure*}

We observe a strong anti-correlation between SSFR$_{H\alpha}$ and $\Delta D$ 
in the sample of all low $D_n4000$ major pair galaxies at $z=0.080-0.376$  
(at all local densities).  The Spearman-rank test computes a probability of no 
correlation of $P_{SR} = 6.0\times10^{-4}$ for the 134 galaxies in the
sample (Figure~\ref{hasepr1}, top left panel.)  The range of absolute luminosities 
and local densities in this sample corresponds to that included in typical 
samples at low redshift \citep[e.g.][]{bgk}.

 Figure~\ref{hasepr1} shows the mean SSFR$_{H\alpha}$ as a function of 
$\Delta D$ for low $D_n4000$ galaxies in low density regions in the volume-limited 
major pairs sample (top right panel).  We measure a correlation between SSFR$_{H\alpha}$ 
and $\Delta D$ for the 70 galaxies in this sample, where 
$P_{SR} =1.3 \times 10^{-2}$.  The decreased 
significance compared to the sample of all major pair galaxies results from the 
reduction in sample size.  Excluding the lowest luminosity galaxies from the 
volume-limited sample may also reduce the signal of the interaction because low 
luminosity galaxies are more strongly affected than the more luminous galaxies.

 We measure a weak correlation between $\Delta D$ and $D_n4000$ for the low
$D_n4000$ galaxies in the volume-limited pair sample at the level of
$P_{SR} = 5.1\times 10^{-2}$; bursts with small $\Delta D$ have small 
$D_n4000$ (Figure~\ref{hasepr1}, bottom left panel).  This trend is consistent 
with the expectation that close pairs have had the most recent starbursts, and
 hence harbor the youngest stellar populations. 

We apply another measure of the recent star formation based on the
stellar population models.  The model-determined fraction 
of total luminosity at  5500~\AA\ from the youngest discrete starburst population age
(5~Myr) correlates with $\Delta D$ for the volume-limited major pair sample:
$P_{SR} = 1.4\times10^{-2}$. 
Pairs at the smallest $\Delta D$ have the highest fractional contribution 
from a 5~Myr stellar population, compared to galaxies with larger $\Delta D$
(Figure~\ref{hasepr1}, bottom right panel).

These results are consistent with other recent observations.
At redshifts of $0.1 <z<1.1$, \citet{lin07} likewise observe an
anti-correlation between the median infrared luminosity of merging galaxies
and pair separation in their sample of $\sim 100$ systems with data from the
DEEP2 Galaxy Redshift Survey and HST/ACS imaging.
 \citet{deravel08} similarly find that the galaxy pairs
at the smallest separations have the greatest median EW([O II]) in their
sample of 251 pairs in the VIMOS VLT Deep Survey \citep{lefevre05},
which has a mean of redshift of $z=0.76$.

We next explore properties of the galaxies that drive the correlation between
star formation indicators and $\Delta D$ --- the galaxies with significant 
star formation rates at small separation.  The galaxies with $\Delta D < 25$~kpc 
and SSFR$_{H\alpha} >4$~M$_{\sun}$~yr$^{-1} / 10^{10}$~M$_{\sun}$ have
luminosities similar to the rest of the low $D_n4000$ galaxies in the 
  volume-limited sample ($M_R \simeq -21.4$).  These 
galaxies are bright; $M_{R}^{*} \simeq -22.1$ for all galaxies in our 
spectroscopic sample in this redshift range (Diaferio et al. 2010, in prep.).
 The correlation between
SSFR$_{H\alpha}$ and $\Delta D$ for luminous galaxies at redshifts
$z=0.1-0.3$ is an  interesting observation for comparison with
lower redshift results.

In the low redshift CfA2 Redshift survey, the 
lower mass galaxies (measured by rotation velocities) exhibit a anti-correlation
 between EW(H$\alpha$) and $\Delta D$;  more massive galaxies do not
(\citet{barton-thesis}).  They conclude that the least massive galaxies exhibit 
the strongest bursts of star formation and are responsible for driving their
observed EW(H$\alpha$)-$\Delta D$ correlation.  
Both \citet{woods07} and \citet{ellison08} find at low redshift that low mass 
(luminosity) galaxies exhibit relatively more powerful triggered specific star formation 
than high mass galaxies.

Finding luminous galaxies with strong evidence of triggered star formation
at intermediate redshift is consistent with  ``downsizing'' \citep{cowie96}, 
which suggests that higher mass (luminosity) galaxies formed their stars earlier,
 and that the lower mass  galaxies have more efficient star formation
at later times than the higher mass galaxies 
\citep{guzman97,brinch00,kodama04,bell05,juneau05,noeske07}.

\subsection{Frequency and strength of triggered star formation} \label{frequency}

Gravitational interactions clearly trigger star formation in some cases (see references 
in \S\ref{intro}).   Some galaxies fail to have enhanced star formation because the pairs have not 
yet reached perigalacticon, the pairs may be mere interlopers along the line-of-sight, 
or they may have internal structure less conducive to gaseous inflows \citep{mihos+hern96}.  
 
\begin{figure*}[htb]
  \begin{center}
 \includegraphics[width=2.25in,angle=90]{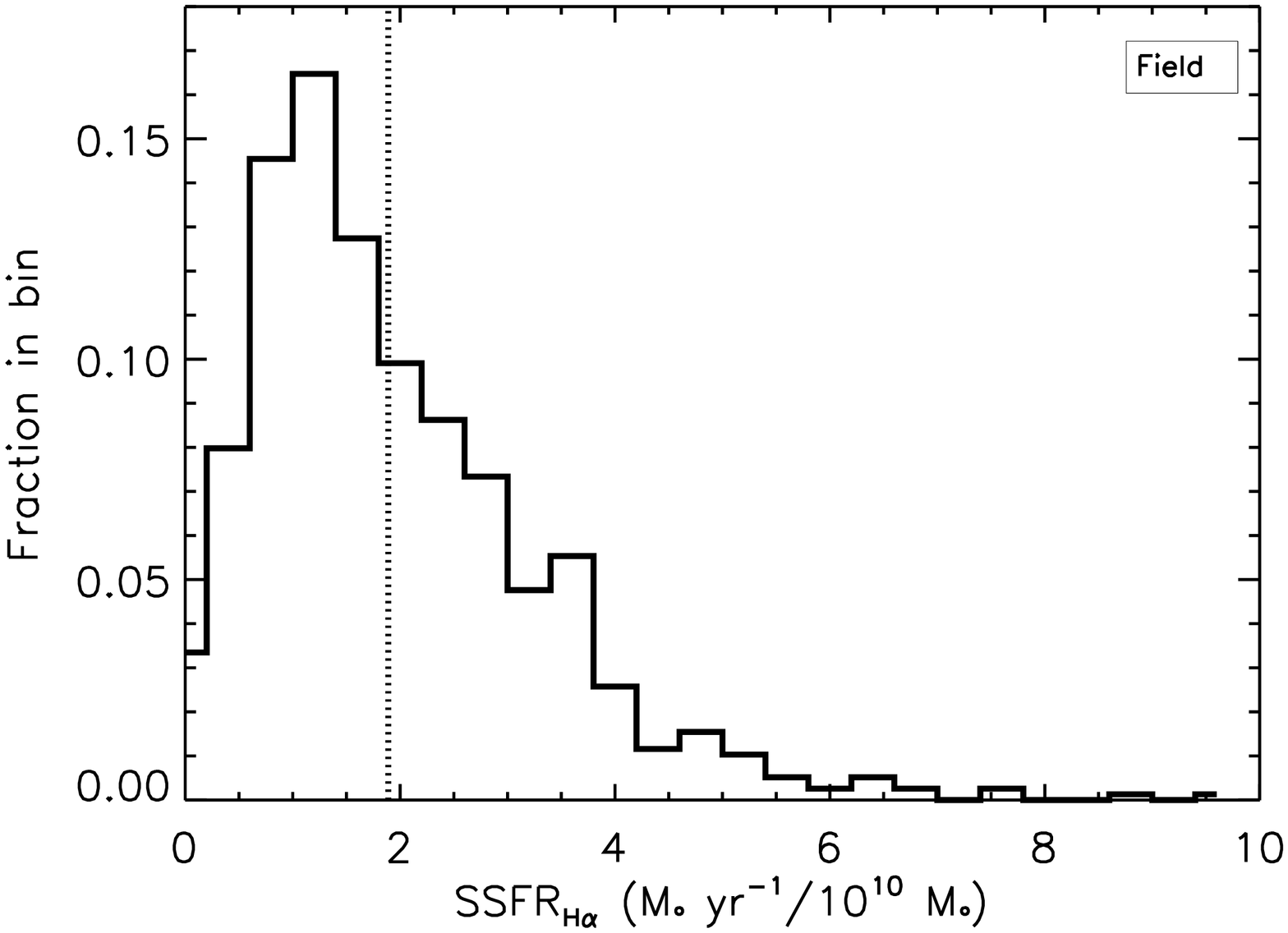}
 \includegraphics[width=2.25in,angle=90]{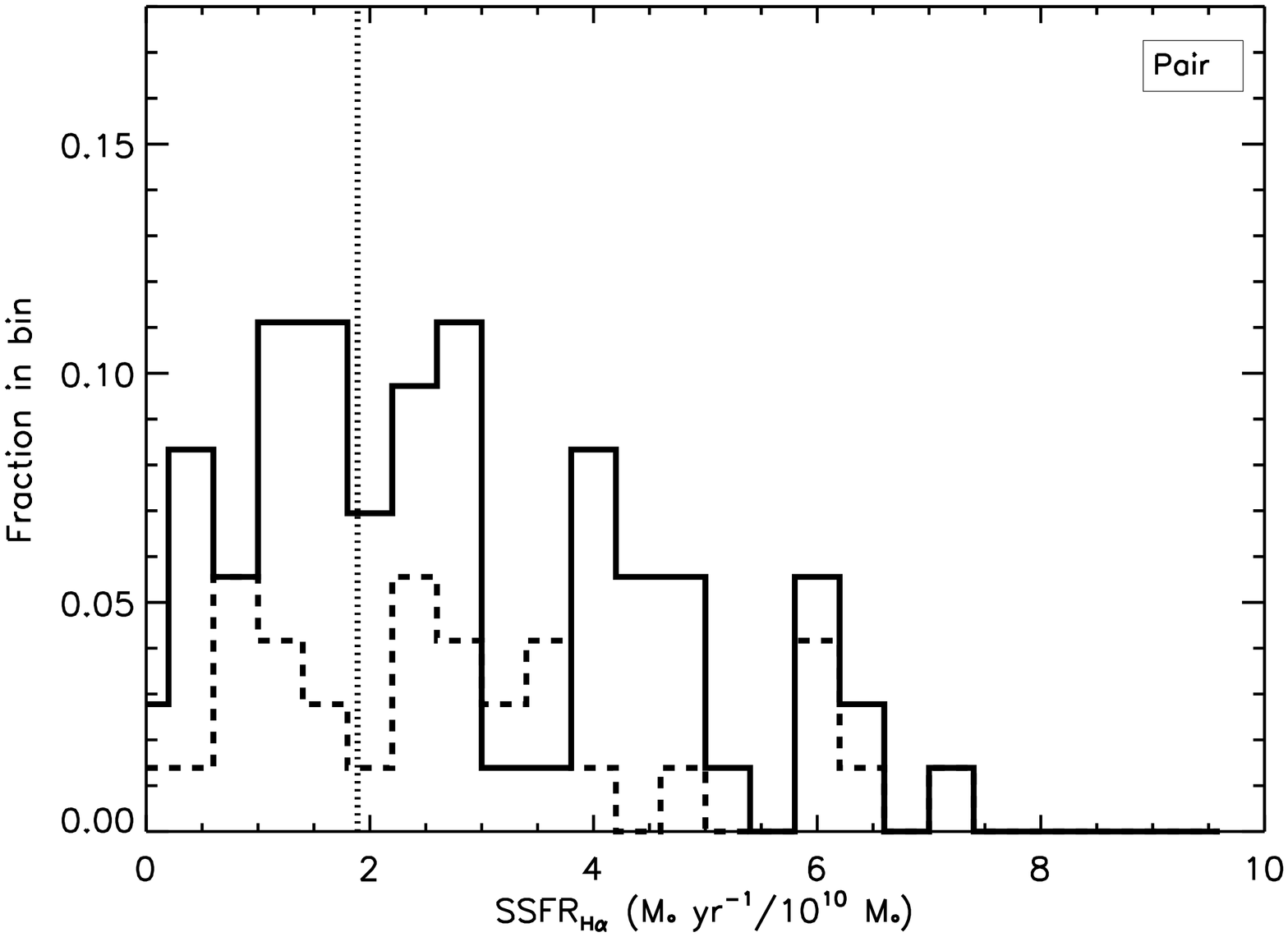}
 \caption{Normalized distribution of SSFR$_{H\alpha}$ for the low $D_n4000$ 
pair ($N = 72$) and field galaxies ($N = 777$) in the volume-limited sample. 
 The pair galaxies' 
distributions exhibit a clear excess of  high SSFR$_{H\alpha}$ compared to that of
the field galaxies.  The dotted line on both plots indicates the median 
SSFR$_{H\alpha}$ for the field galaxies. The dashed line on the pair galaxies'
plot shows the contribution from galaxies with $\Delta D < 25$~kpc.}
 \label{dist_sfr3}
 \end{center}
\end{figure*}

We place limits on the frequency of triggered star formation by comparing the 
SSFR$_{H\alpha}$ 
of the low $D_n4000$ pair galaxies with those of the low $D_n4000$ unpaired field 
galaxies in the volume-limited samples.  Figure~\ref{dist_sfr3} shows the normalized 
distributions of SSFR$_{H\alpha}$ for the low $D_n4000$ galaxies in major pairs and the 
unpaired field galaxies. The dotted line indicates the median SSFR$_{H\alpha}$ 
for the field galaxies.
 Table~\ref{fracmedian} lists the quartiles of the distribution and the
fraction of pair galaxies with SSFR$_{H\alpha}$ more than twice
 the field median.  The results for the very close pairs, 
$\Delta D < 25$~kpc, are listed separately.

\begin{deluxetable}{llllll}
\tablecaption{Distribution of specific star formation rates in pairs and unpaired field galaxies
 \label{fracmedian} }
\tablewidth{0pc} 
\tablehead{
\colhead{Sample\tablenotemark{a}} & \colhead{$Q_{25}$\tablenotemark{b}} & \colhead{$Q_{50}$} & 
\colhead{$Q_{75}$} & \colhead{$> 2 F_{50}$\tablenotemark{c}} &  \colhead{Sample size}}
\startdata
\colhead{Pair}                        & 1.55  & 2.61  & 4.26  & $32\pm 7\%$  & 72 \\
\colhead{Close pair\tablenotemark{d}} & 1.87  & 3.11  & 4.43  & $42\pm 13\%$ & 26 \\
\colhead{Field}                       & 1.18  & 1.80  & 2.86  & $14\pm 1\%$  & 777 \\
\enddata
\tablenotetext{a}{Pair and field samples derive from the volume-limited sample
of low $D_n4000$ galaxies, where  $M_R < -20.8$.  Pair galaxies are
 in major pairs, $\left | \Delta M_R \right | < 1.75$.}
\tablenotetext{b}{Quartile of the distribution of SSFR$_{H\alpha}$ 
in units of $M_{\sun} yr^{-1}/10^{10} M_{\sun}$.}
\tablenotetext{c}{Fraction of pair galaxies with SSFR$_{H\alpha} > 2 F_{50}$, 
where $F_{50}$ is the median of the field galaxies in the same redshift bin.}
\tablenotetext{d}{Close pair: $\Delta D < 25$~kpc.}
\end{deluxetable}

Table~\ref{fracmedian} shows that the  low $D_n4000$ major pair galaxies
have an excess of high SSFR$_{H\alpha}$ compared to the low $D_n4000$ field galaxies.
We define the fraction of pair galaxies  experiencing enhanced specific star formation 
as SSFR$_{H\alpha} =2 \times F_{50}$,  where $F_{50}$ is the 
median SSFR$_{H\alpha}$ for the field galaxies.  According to this definition, 
$32\pm 7\%$  of major pair galaxies experience enhanced
specific star formation.   A greater fraction of close pairs  with
$\Delta D < 25$~kpc, exhibit enhanced specific star formation rates: $42\pm 13\%$.
  Observing a greater fraction of close pairs with enhanced star formation
is consistent with triggering at the closest approach and the subsequent 
decline as the burst ages and the pair moves apart.   We find only one galaxy with 
 SSFR$_{H\alpha} > 5\times F_{50}$.

A number of generic selection issues affect the measured frequency and strength of 
triggered star formation.  The first issue is that a small fraction ($< 4\%$)
of the field galaxies is likely to be in a pair with an undetected companion (\S\ref{fieldsample}).
 Second, excluding galaxies with strong AGN preferentially excludes pairs with
active star formation  because triggered AGN activity and star formation 
activity often occur as part of the same process 
(\citealp[e.g.][]{hopkins08_356}).  Third,
we are unable to resolve some pairs or mergers at separations $\Delta D <15$~kpc,
where the strongest star formation enhancement is expected \citep[e.g.][]{bgk}.
Our measurement of the frequency and strength of triggered star formation 
should therefore be interpreted as lower limits.

We compare measurement of the frequency of triggered star formation
 from our volume-limited major pairs sample with
the predictions of \citet{dimatteo08}, who use numerical simulations to study the 
frequency, intensity, and duration of triggered star formation activity.  
Between 25 and $50\%$ of their ``fly-bys'' have star formation rates
twice the isolated case (see their Table D.1.)  We compare with their
fly-bys and not their merger scenario because the galaxies in our sample
 are distinct systems, which have not yet had a final merger. 
Our results are consistent with the predictions of Di Matteo et al. 

One difference between our observations and the Di Matteo et al.
simulations is that their maximum star formation rate  refers to the lifetime of 
the galaxy; ours represents a snapshot in time.  Because the most
intense bursts of star formation  occur over relatively short time
scales in the course of the merger (see Figure 4 in Di Matteo et al.),
we are unlikely to observe pair galaxies at maximum intensity.
We therefore expect to measure a  lower frequency than that
predicted by Di Matteo et al.  This prediction is consistent with our
finding only one galaxy with  SSFR$_{H\alpha} > 5\times F_{50}$;
Di Matteo et al. find that $15\%$ of their major mergers have 
star formation enhanced by this large factor.

The frequency and strength of triggered star formation that we measure
are consistent with the observations of \citet{jogee08}, who find that the average 
star formation rate of strongly disturbed galaxies exhibits only a modest 
increase over the morphologically undisturbed galaxies in their sample of 
$\sim4,500$ galaxies at $0.24<z<0.80$.  Our results are also in line 
with the low redshift observations of \citet{ellison08}, who measure 
star formation rate enhancement in SDSS galaxies at $z<0.16$. \citet{li08} find 
that the SDSS systems with the highest star formation rates are likely to have 
a companion, but not all systems with close companions have high star
formation rates.   

A consistent explanation for the range of results drawn from simulations
and observations is that strongly enhanced star formation is rare and short 
lived.  Simulations can track the maximum enhancement across the lifetime of the
merger; different types of pair selection probes systems in different stages
of the interaction.  Density effects may also reduce the impact of observations
of star formation in pairs \citep{barton07}.

\subsection{Duration of triggered star formation} \label{models}

We use the stellar population models described in \S\ref{sfr}
 to compare the stellar composition of pair and unpaired field galaxies.
The stellar population models fit the spectra with a discrete set of starbursts 
of age 0.005, 0.025, 0.1, 0.3, 0.6, 0.9, 1.4, 2.5, 5 and 10~Gyr.  
From the models we extract the contribution of each starburst population to the flux at 
5500~\AA. 

\begin{figure}[htb]
  \begin{center}
 \includegraphics[width=2.25in,angle=90]{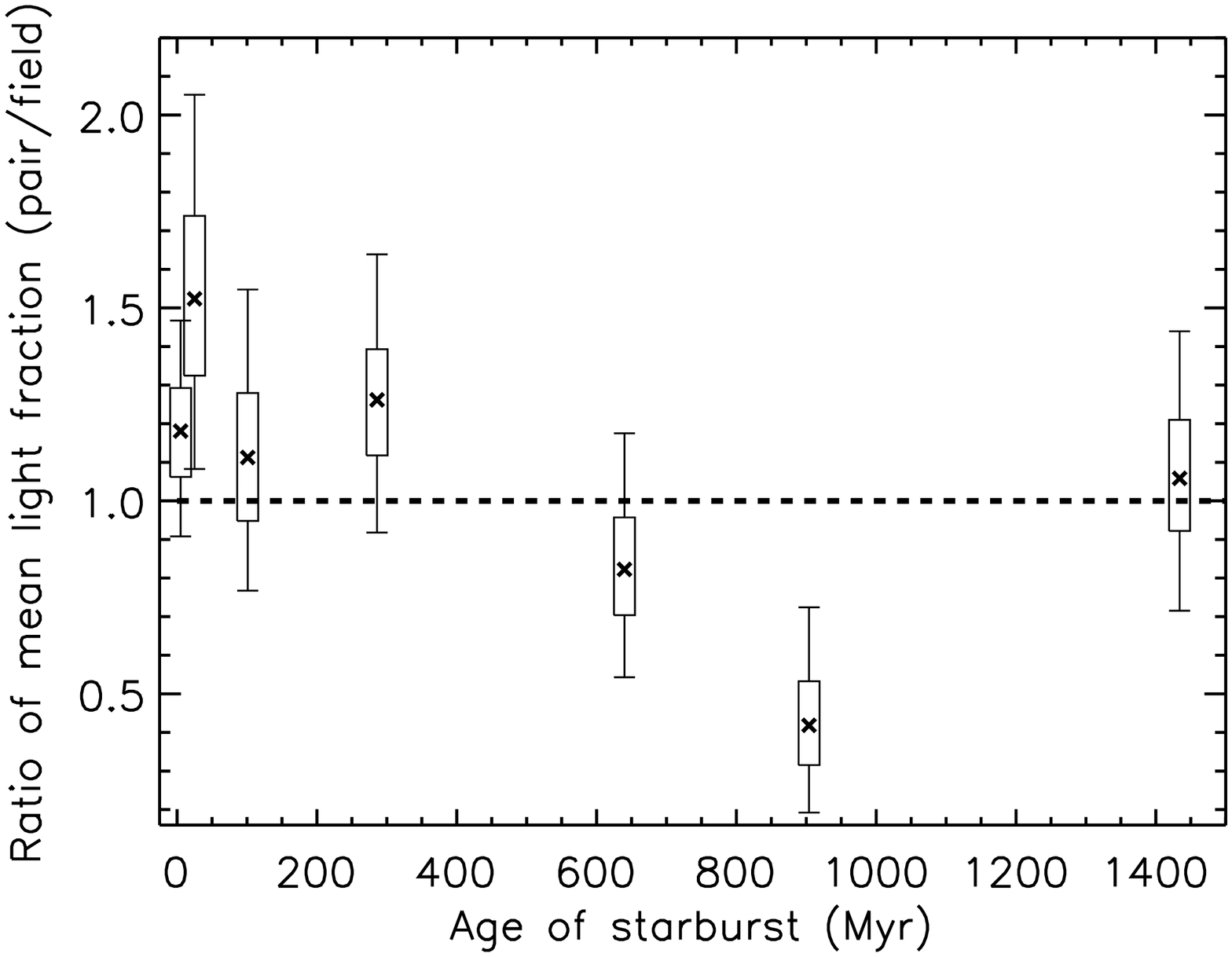}
 \caption{Ratio of mean light fraction from starburst populations as a function of 
starburst population age for pair and unpaired field  galaxies with low 
$D_n4000$ in the volume-limited sample. The pair
galaxies show a greater light fraction from young stellar populations
up to burst ages $\sim 300-400$~Myr.  The dashed line indicates a ratio of unity.
Error bars are from bootstrap re-sampling: boxes indicate the inter-quartile range
and the outer lines show the $95\%$ confidence interval.}
 \label{fluxratio}
 \end{center}
\end{figure}

Figure \ref{fluxratio} shows the ratio (pair/field) of the mean fraction of 
flux at 5500~\AA~attributed to each discrete starburst as a function of starburst 
age for the starburst populations included in the model.  The pair galaxies 
clearly contain a larger fraction of young stellar populations up to burst 
ages $\sim 300-400$~Myr.  The ratio of the mean fraction of flux from each 
starburst population dips slightly below one for burst ages $\gtrsim 500$~Myr,
because the younger stellar populations contain a greater fraction of the flux
 in the pair galaxies.

Our measurement agrees well with the results of
\citet{bgk}, who  apply spectral synthesis models \citep{l99,bruzual+ch96} 
to their data from the CfA2 Redshift Survey \citep{cfa2}.  They 
determine that the H$\alpha$ emission and the galaxy colors are best 
described by a burst with a continuous duration $\gtrsim 100$~Myr on top of the 
pre-existing stellar population.  The simulations of \citet{dimatteo08} 
similarly suggest a merger-driven starburst duration of up to a few hundred Myr,
 consistent with our measured duration.

\section{Summary and conclusion} \label{conclusion}

We examine spectroscopic properties of pair and field galaxies to quantify effects 
of gravitational interactions at intermediate redshift.  Our  sample derives from 
the Smithsonian Hectospec Lensing Survey \citep[SHELS;][]{geller05,geller10}.
 SHELS includes 9,825 galaxies and is $97.7\%$ spectroscopically 
complete to $R=20.3$ over an area of 4~deg$^2$.  We select for galaxies in the 
redshift range $z=0.080-0.376$. This survey represents the most complete 
spectroscopic sample in its redshift range.

We focus on the systems that have the potential to exhibit bursts of star formation
as a result of the interaction.  Substantial evidence shows that major pairs
are more strongly affected by the interaction.  We identify a full set of major 
($\left | \Delta M_R \right | < 1.75$) pairs including 622 galaxies in the 
redshift range $z = 0.080-0.376$,  and a volume-limited subset of major pairs 
in the redshift range $z = 0.080-0.310$, including 327 galaxies to $M_R = -20.8$.

Within our major pair sample, we further narrow our selection using the spectroscopic 
index $D_n4000 = 1.44$ as the divide between systems with older stellar populations, and 
systems with young stellar populations that likely contain gas. 
We further restrict our sample to systems with low surrounding density, which we
measure with a count of neighbors within a volume of comoving radius 985~kpc.

The spectroscopic diagnostics of  H$\alpha$ specific
star formation rate (SSFR$_{H\alpha}$), $D_n4000$, and a set of stellar population 
models  enable the investigation of the strength, frequency, and timescale of 
triggered star formation.   We show:

\begin{enumerate}
\item The spectroscopic indicator $D_n4000$ provides a useful 
classification metric and corresponds closely with identification of
emission line galaxies.
\item The star formation  indicators SSFR$_{H\alpha}$, $D_n4000$, and presence of young 
stellar populations exhibit an anti-correlation with $\Delta D$, demonstrating that 
bursts of star formation are associated with close proximity to a major 
companion.
\item  $32\pm 7\%$ of major pair galaxies in the volume-limited sample 
experience enhanced specific star formation activity at
twice the median of the unpaired field galaxies.
For very close pairs ($\Delta D< 25$~kpc), the fraction is $42\pm 13\%$.
This trend is consistent with the tidal triggering picture.
\item  We use stellar population models to show the burst of star formation following 
an interaction has a duration of $\sim300-400$~Myr.  Pair galaxies show an increase over 
the field in the light fraction from young stellar populations for burst ages up
 to $\sim 300-400$~Myr. 
\end{enumerate}

The most effective way to  increase our ability to measure differences between 
pair and field galaxies as a function of redshift, or to determine the AGN fraction in 
pairs would be to observe a larger population of very close pairs ($\Delta D < 15$~kpc) 
at redshift $z\sim 0.3$ using small aperture spectroscopy.  It is important to have 
high resolution photometric data in combination with good seeing to distinguish 
close pairs.

\section*{Acknowledgment}

This work benefited greatly from discussions with 
Elizabeth Barton, Nelson Caldwell, Scott Kenyon, and Lisa Kewley.
We thank the members of DFW's PhD thesis committee for their
comments that improved this work: 
Lars Hernquish, Robert Kirshner, and Andrew Szentgyorgyi.
We thank the anonymous referee for a helpful and knowledgeable
report.

We gratefully acknowledge the contribution of the CfA's Telescope 
Data Center team, especially Doug Mink, Susan Tokarz, and William Wyatt 
for their work with the Hectospec data reduction pipeline.  We thank
 the Hectospec engineering team, including Robert Fata,
Tom Gauron, Edward Hertz, Mark Mueller, and Mark Lacasse, and the
instrument specialists Perry Berlind and Michael Calkins, along
with the rest of the staff at the MMT Observatory.

Funding for the SDSS and SDSS-II has been provided by the Alfred P. 
Sloan Foundation, the Participating Institutions, the National Science 
Foundation, the U.S. Department of Energy, the National Aeronautics and
 Space Administration, the Japanese Monbukagakusho, the Max Planck 
Society, and the Higher Education Funding Council for England. 
The SDSS Web Site is http://www.sdss.org/.

{\it Facilities:} \facility{MMT (Hectospec)}, \facility{Sloan}, \facility{KPNO:CFT (Mosaic)}

\end{document}